\pdfoutput=1
\documentclass[10pt,twocolumn,twoside]{IEEEtran}
\IEEEoverridecommandlockouts
\usepackage{amsmath,graphicx,algorithmic,algorithm,amssymb,mathdots,setspace,color}
\usepackage{array}
\usepackage{amsfonts}
\usepackage{arydshln}
\usepackage{balance}
\usepackage{hyperref}
\DeclareMathOperator*{\argmin}{argmin}

\newtheorem{theorem}{\textbf{Theorem}}

\renewcommand{\l}{\ell}
\newcommand{\norm}[1]{\lVert#1\rVert}
\newcommand{\card}[1]{\lvert#1\rvert}

\newcommand\Y{{\mathcal{Y}}}

\newcommand\Acal{\mathcal{A}}
\newcommand\Bcal{\mathcal{B}}
\newcommand\T{\mathcal{T}}
\renewcommand\S{\mathcal{S}}
\newcommand{\N}{{\mathcal{N}}}

\newcommand{\E}{{\mathcal{E}}}
\newcommand{\K}{{\mathcal{K}}}
\newcommand{\Hcal}{{\mathcal{H}}}
\newcommand{\G}{{\mathcal{G}}}
\newcommand{\D}{{\mathcal{D}}}
\newcommand{\V}{{\mathcal{V}}}

\providecommand{\card}[1]{\lvert#1\rvert}

\newcommand{\Rbb}{\mathbb{R}}

\renewcommand{\L}{{\mathcal{L}}}

\newcommand{\transpose}{{\!\scriptscriptstyle\mathrm T}}  

\def\L{{\cal L}}

\DeclareMathOperator*{\argmax}{arg\,max}

\title{A Multiscale Pyramid Transform for Graph Signals}
\author{\IEEEauthorblockN{David I Shuman, Mohammad Javad Faraji, and Pierre Vandergheynst}
\thanks{This work was supported in part by FET-Open grant number 255931 UNLocX.} 
\thanks{Part of this work appeared in the master's thesis \cite{faraji_thesis} and was presented at the Duke Workshop on Sensing and Analysis of High-Dimensional Data in July 2011. Slides and video are available at 
http://sahd.pratt.duke.edu/2011\textunderscore files/Videos\textunderscore and\textunderscore Slides.html
} 
\thanks{David I Shuman is with the Department of Mathematics, Statistics, and Computer Science, Macalester College, St. Paul, MN 55105, USA (email: dshuman1@macalester.edu). 
Mohammad Javad Faraji is with the Computational Neuroscience Laboratory (LCN), Ecole Polytechnique F{\'e}d{\'e}rale de Lausanne (EPFL), School of Computer and Communication Sciences and Brain Mind Institute, School of Life Sciences, CH-1015 Lausanne, Switzerland (email: mohammadjavad.faraji@epfl.ch). 
Pierre Vandergheynst is with the Signal Processing Laboratory (LTS2), Ecole Polytechnique F{\'e}d{\'e}rale de Lausanne (EPFL), Institue of Electrical Engineering, CH-1015 Lausanne, Switzerland (email: pierre.vandergheynst@epfl.ch).  }
\thanks{The authors would like to thank the anonymous reviewers for their many constructive comments on earlier versions of this paper.}
\thanks{A MATLAB implementation of the pyramid transform proposed in this paper is available in the Graph Signal Processing Toolbox \cite{gspbox}, which is publicly available at \url{https://lts2.epfl.ch/gsp/}. MATLAB code for the numerical experiments in Section VI is available at \url{http://www.macalester.edu/\textasciitilde dshuman1/publications.html}}
}

\begin{document}
%
\maketitle

\begin{abstract}
Multiscale transforms designed to process analog and discrete-time signals and images cannot be directly applied to analyze high-dimensional data residing on the vertices of a weighted graph, as they do not capture the intrinsic 
topology of the 
graph data domain. 
In this paper, we 
adapt the Laplacian pyramid transform for signals on Euclidean domains so that it can 
be used to analyze high-dimensional data residing on the vertices of a weighted graph. Our approach is to study existing methods and develop new methods for the
four  
fundamental operations of graph downsampling, graph reduction, and filtering and interpolation of  
signals on graphs. 
Equipped with appropriate notions of these operations, we leverage the basic multiscale constructs and intuitions from classical signal processing to generate a transform that yields both a multiresolution of graphs and an associated multiresolution of a graph signal on the underlying sequence of graphs.
\end{abstract}
\section{Introduction}\label{sec:intro}
Multiscale 
transform methods can reveal structural information about signals, such as singularities or irregular structural patterns, at different resolution levels. At the same time, via coarse-to-fine analysis, they provide a way to reduce the complexity and dimensionality of many signal processing tasks, often resulting in fast algorithms.
However, multiscale transforms such as wavelets and filter banks designed to process analog and discrete-time signals and images cannot be directly applied to analyze high-dimensional data residing on the vertices of a weighted graph, as they do not capture the intrinsic 
topology
of the underlying graph data domain (see \cite{shuman_SPM} for an overview of the main challenges of this emerging field of signal processing on graphs).

To address this issue, classical wavelets have recently been generalized to the graph setting in a number of different ways. Reference \cite{shuman_SPM} contains a more thorough review of these graph wavelet constructions, which include, e.g.,  spatially-designed graph wavelets \cite{Crovella2003}, diffusion wavelets \cite{diffusion_wavelets}, spectral graph wavelets \cite{sgwt}, lifting based wavelets \cite{jansen,narang_lifting_graphs} critically sampled two-channel wavelet filter banks \cite{narang_icip}, critically-sampled spline wavelet filter banks \cite{ekambaram_globalsip}, and multiscale wavelets on balanced trees \cite{gavish}. Multiresolutions of graphs also have a long history in computational science problems including graph clustering, numerical solvers for linear systems of equations (often arising from discretized differential equations), combinatorial optimization problems, and computational geometry (see, e.g., \cite{teng}\nocite{brandt_new_survey}\nocite{ron}-\cite{vishnoi} and references therein). We discuss some of the related work from these fields in Sections \ref{Se:alt_down} and \ref{Se:alternatives}.

In this paper, we present a modular framework 
for adapting Burt and Adelson's Laplacian pyramid transform \cite{burt_adelson} to the graph setting.
Our main contributions are to (1) survey different methods for and desirable properties of
the four fundamental graph signal processing operations of graph downsampling,  graph reduction, generalized filtering, and interpolation of graph signals 
 (Sections \ref{Se:down}-\ref{Se:filtering});
 (2) present  
 new graph downsampling and reduction methods, including downsampling based on the polarity of the largest Laplacian eigenvector and Kron reduction followed by spectral sparsification;
 and (3) leverage these fundamental operations to construct a new multiscale pyramid transform that yields both a multiresolution of a graph and a multi-scale analysis of a signal residing on that graph 
 (Section \ref{Se:pyramid}).
 We also discuss some implementation approximations and open issues in Section \ref{Se:approximations}, as it is important that the computational complexity of the resulting multiscale transform scales well with the number of vertices and edges in the underlying graph.

\section{Spectral Graph Theory Notation}
We consider connected, loopless (no edge connecting a vertex to itself), undirected, weighted graphs. We represent such a graph by the triplet $\G=\{\V,\E,w\}$, where $\V$ is a set of $N$ vertices, $\E$ is a set of edges, and $w:\E \rightarrow \Rbb^+$ is a weight function that assigns a non-negative weight to each edge. An equivalent representation is $\G=\{\V,\E,\mathbf{W}\}$, where
$\mathbf{W}$ is a $N \times N$ weighted adjacency matrix with nonnegative entries
\begin{align*}
{W}_{ij}=\begin{cases}
 w(e),&\mbox{if }e \in \E\mbox{ connects vertices }i\mbox{ and }j\\
 0,&\mbox{if no edge connects vertices }i\mbox{ and }j
\end{cases}.
\end{align*}
In unweighted graphs, the entries of the adjacency matrix $\mathbf{W}$ are ones and zeros, with a one corresponding to an edge between two vertices and a zero corresponding to no edge. The degree matrix $\mathbf{D}$ is a diagonal matrix with an $i^{th}$ diagonal element  $D_{ii}=d_i=\sum_{j \in \N_i}W_{ij}$, where $\N_i$ is the set of vertex $i$'s neighbors in $\G$. Its maximum element is $d_{\max}:=\max_{i \in \V} \{d_i\}$. We denote the combinatorial graph Laplacian by $\L:=\mathbf{D}-\mathbf{W}$, the normalized graph Laplacian by $\tilde{\L}:=\mathbf{D}^{-\frac{1}{2}}\L \mathbf{D}^{-\frac{1}{2}}$, and their respective eigenvalue and eigenvector pairs by $\left\{(\lambda_{\l},\mathbf{u}_{\l})\right\}_{\l=0,1,\ldots,N-1}$ 
and $\{(\tilde{\lambda}_{\l},\tilde{\mathbf{u}}_{\l})\}_{\l=0,1,\ldots,N-1}$. Then $\mathbf{U}$ and $\tilde{\mathbf{U}}$ are the matrices whose columns are equal to the eigenvectors of $\L$ and $\tilde{\L}$, respectively. We assume without loss of generality that the eigenvalues are monotonically ordered so that $0=\lambda_{0}<\lambda_{1}\leq \lambda_{2} \leq \ldots \leq \lambda_{N-1}$, and we denote the maximum eigenvalues and associated eigenvectors by $\lambda_{\max}=\lambda_{N-1}$ and $\mathbf{u}_{\max}=\mathbf{u}_{N-1}$. The maximum eigenvalue $\lambda_{\max}$ is said to be \emph{simple} if $\lambda_{N-1}>\lambda_{N-2}$.

\section{Graph Downsampling} \label{Se:down}
Two key components of multiscale transforms for discrete-time signals are downsampling and upsampling.\footnote{We focus here on downsampling, as we are only interested in upsampling previously downsampled graphs. As long as we track the positions of the removed components of the signal, it is straightforward to upsample by inserting zeros back into those components of the signal.} To downsample a discrete-time sample by a factor of two, we remove every other component of the signal. 
To extend many ideas from classical signal processing to the graph setting, we need to define a notion of downsampling for signals on graphs. Yet, it is not at all obvious what it means to remove \emph{every other} component of a signal $\mathbf{f} \in \Rbb^N$ defined on the vertices of a graph. In this section, we outline desired properties of a downsampling operator for graphs, and then go on to suggest one particular downsampling method.

Let $\D: \G=\{\V,\E,\mathbf{W}\} \rightarrow 2^{\V}$ be a graph downsampling operator that maps a weighted, undirected graph to a subset of vertices $\V_1$ to keep. The complement $\V_1^c := \V \backslash \V_1 =\left\{v \in \V: v\notin \V_1 \right\}$ is the set of vertices that $\D$ removes from $\V$. Ideally, we would like the graph downsampling operator $\D$ to have the following properties:
\begin{enumerate}
\item[(D1)]  It removes approximately half of the vertices of the graph (or, equivalently, approximately half of the components of a signal on the vertices of the graph); i.e., $\card{\D(\G)} =\card{\V_1} \approx \frac{\card{\V}}{2}$.
\item[(D2)] The set of removed vertices are not connected with edges of high weight, and the set of kept vertices are not connected with edges of high weight; i.e., if $i,j \in \V_1$, then ${W}_{ij}$ is low, and if $i,j \in \V_1^c$, then ${W}_{ij}$ is low.
\item[(D3)] It has a computationally efficient implementation.
\end{enumerate}

\subsection{Vertex Selection Using the Largest Eigenvector of the Graph Laplacian} \label{Se:largest_eigenvector}
The method we suggest to use for graph downsampling 
is to select the vertices to keep based on the polarity of the components of the largest eigenvector; namely, let
\begin{align}\label{Eq:largest_eig_method}
\V_1=\V_+:=\left\{i \in \V: {{u}}_{\max}(i) \geq 0 \right\}. 
\end{align}
We refer to this method as the largest eigenvector vertex selection method.
A few remarks are in order regarding this choice of downsampling operator. First, the polarity of the largest eigenvector splits the graph into two components. In this paper, we choose to keep the vertices in $\V_+$, and eliminate the vertices in $\V_-:=\left\{i \in \V: {{u}}_{\max}(i) < 0 \right\}$, but we could just as easily do the reverse, or keep the vertices in $\V_{big}:=\argmax_{\V_1 \in \left\{\V_+,\V_-\right\}}\card{\V_1}$, for example. Second, for some graphs such as the complete graph, $\lambda_{\max}$ is a repeated eigenvalue, so the polarity of  ${\mathbf{u}}_{\max}$ is not uniquely defined. In this case, we arbitrarily choose an eigenvector from the eigenspace. Third, we could just as easily base the vertex selection on the polarity of the normalized graph Laplacian eigenvector, $\tilde{\mathbf{u}}_{\max}$ associated with the largest eigenvalue, $\tilde{\lambda}_{\max}$. In some cases, such as the bipartite graphs discussed next, doing so yields exactly the same selection of vertices as downsampling based on the largest non-normalized graph Laplacian eigenvector; however, this is not true in general.

In the following sections, we motivate the use of the largest eigenvector of the graph Laplacian from two different perspectives - first from a more intuitive view as a generalization of downsampling techniques for special types of graphs, and then from a more theoretical point of view by connecting the vertex selection problem to graph coloring, spectral clustering, and nodal domain theory.

\subsection{Special Case: Bipartite Graphs} \label{Se:vs_special}
There is one situation in which there exists a fairly clear notion of removing every other component of a graph signal -- when the underlying graph is 
\emph{bipartite}.  A graph $\G = \{\V,\E,\mathbf{W}\}$ is \emph{bipartite} if the set of vertices $\V$ can be partitioned into two subsets ${\V}_1$ and ${\V}_1^c$ so that every edge $e \in \E$ links one vertex in ${\V}_1$ with one vertex in ${\V}_1^c$. In this case, it is natural to downsample by keeping all of the vertices in one of the subsets, and eliminating all of the vertices in the other subset. In fact, as stated in the following theorem, the largest eigenvector downsampling method does precisely this in the case of bipartite graphs.

\begin{theorem}[Roth, 1989] \label{Th:bipartite}
For a connected, bipartite graph $\G = \{\V_1 \cup \V_1^c,\E,\mathbf{W}\}$, the largest eigenvalues, $\lambda_{\max}$ and $\tilde{\lambda}_{\max}$, of $\L$ and $\tilde{\L}$, respectively, are simple, and $\tilde{\lambda}_{\max}=2$. Moreover, the polarity of the components of the eigenvectors $\mathbf{u}_{\max}$ and $\tilde{\mathbf{u}}_{\max}$ associated with $\lambda_{\max}$ and $\tilde{\lambda}_{\max}$ both split $\V$ into the bipartition $\V_1$ and $\V_1^c$. That is, for $\mathbf{v}=\mathbf{u}_{\max}$ or $\mathbf{v}=\tilde{\mathbf{u}}_{\max}$,
\begin{align}\label{Eq:split}
{v}(i){v}(j) > 0,&\mbox{ if }i,j \in \V_1\mbox{ or }i,j \in \V_1^c, \mbox{ and } \nonumber \\
{v}(i){v}(j) < 0,&\mbox{ if }i \in \V_1, j \in \V_1^c \mbox{ or }i \in \V_1^c ,j \in \V_1.
\end{align}
If, in addition, $\G$ is $k$-regular ($d_i=k,~\forall i\in \V$), then $\lambda_{\max}=2k$, and
\begin{align*}
\mathbf{u}_{\max} = \tilde{\mathbf{u}}_{\max} = \left\{
\begin{array}{rl}
\frac{1}{\sqrt{N}}, &\hbox{if }i \in \V_1 \\
-\frac{1}{\sqrt{N}}, &\hbox{if }i \in \V_1^c
\end{array}
\right. .
\end{align*}
\end{theorem}
The majority of the statements in Theorem \ref{Th:bipartite} follow from results of Roth in \cite{roth}, which are also presented in
\cite[Chapter 3.6]{lap_eigen}. 

\begin{figure}[tb]
\centering
\begin{minipage}[b]{.45 \linewidth}
   \centering
   \centerline{\includegraphics[width=\linewidth]{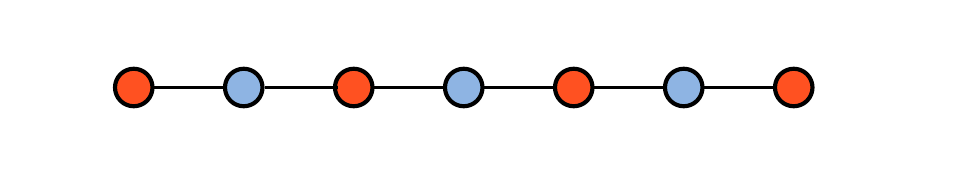}}
\centerline{\small{(a)}}
\end{minipage} 
\begin{minipage}[b]{.45 \linewidth}
   \centering
   \centerline{\includegraphics[width=\linewidth]{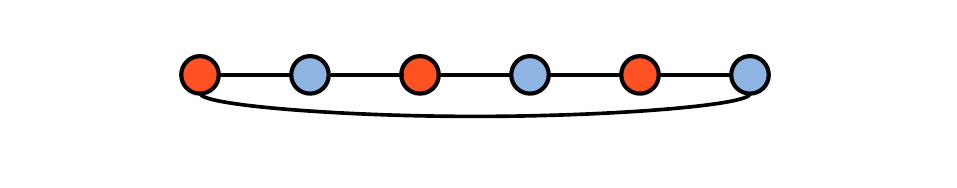}}
\centerline{\small{(b)}}
\end{minipage}\\
\vspace{0.4cm}
\begin{minipage}[b]{.45 \linewidth}
   \centering
   \centerline{\includegraphics[width=\linewidth]{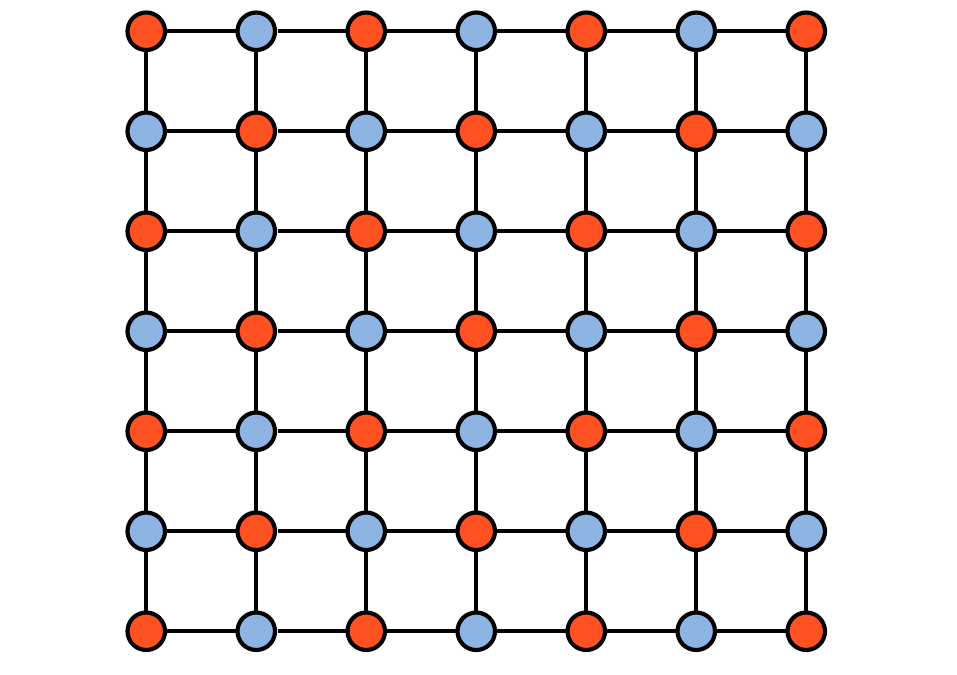}}
\centerline{\small{(c)}}
\end{minipage} 
\begin{minipage}[b]{.45 \linewidth}
   \centering
   \centerline{\includegraphics[width=\linewidth]{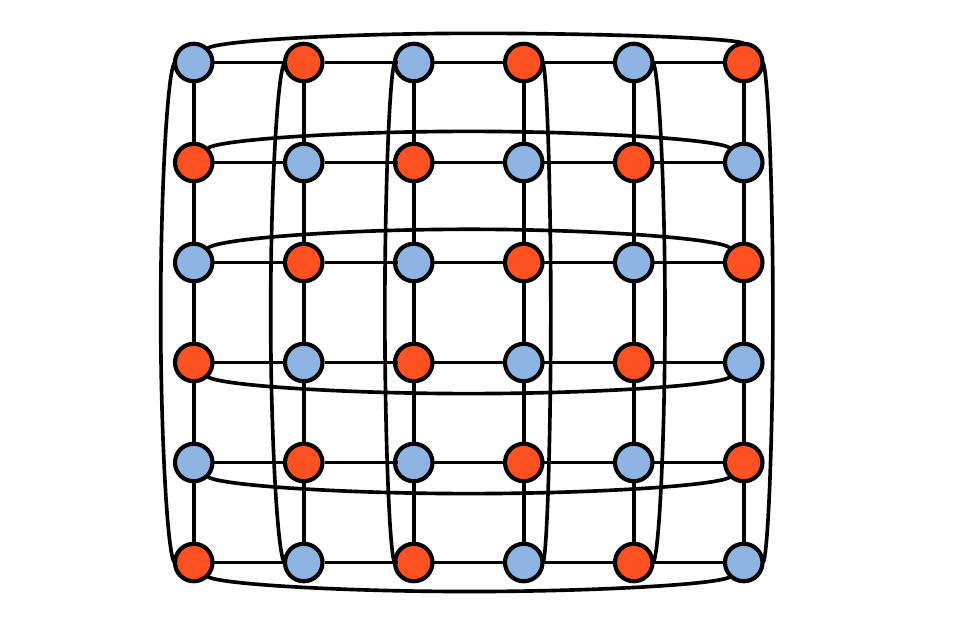}}
\centerline{\small{(d)}}
\end{minipage}\\
\vspace{0.4cm}
\begin{minipage}[b]{\linewidth}
   \centering
   \centerline{\includegraphics[width=.45\linewidth]{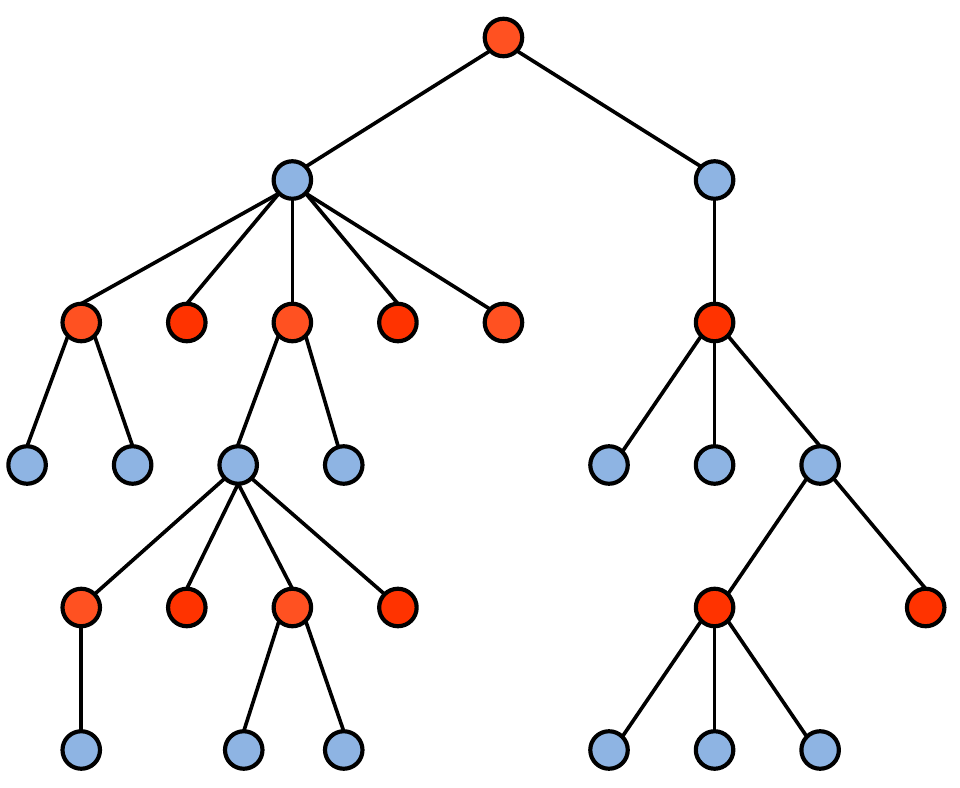}}
\centerline{\small{(e)}}
\end{minipage}
\caption {Examples of partitioning structured graphs into two sets (red and blue) according to the polarity of the largest eigenvector of the graph Laplacian. For the path graph in (a) and the ring graph in (b), the method selects every other vertex like classical downsampling of discrete-time signals. For the finite grid graphs with the ends unconnected (c) and connected (d), the method results in the quincunx sampling pattern. For a tree graph in (e), the method groups the vertices at every other depth of the tree.}
  \label{Fig:special_graphs}
\end{figure}

The path, ring (with an even number of vertices), and finite grid graphs, which are shown in Figure \ref{Fig:special_graphs}, are all examples of bipartite graphs and all have simple largest graph Laplacian eigenvalues. Using the largest eigenvector downsampling method leads to the elimination of every other vertex on the path and ring graphs, and to the quincunx sampling pattern on the finite grid graph (with or without boundary connections). 

Trees (acyclic, connected graphs) are also bipartite. An example of a 
tree is shown in Figure \ref{Fig:special_graphs}(e). Fix an arbitrary vertex $r$ to be the root of the tree $\T$, let $\Y_r^0$ be the singleton set containing the root, and then define the sets $\left\{\Y_r^{t}\right\}_{t=1,2,\ldots}$ by $\Y_r^t:=\left\{i\in \V: i \mbox { is } t \mbox{ hops from the root vertex $r$ in } \T\right\}$. Then the polarity of the components of largest eigenvector of the graph Laplacian splits the vertices of the tree into two sets according to the parity of the depths of the tree. That is, if we let $\Y_r^{even}:=\cup_{t=0,2,\ldots}\Y_r^t$ and $\Y_r^{odd}:=\cup_{t=1,3,\ldots}\Y_r^t$, then $\Y_r^{even}=\V_+$ and $\Y_r^{odd}=\V_-$, or vice versa.


In related work, \cite{narang_down} and \cite{narang_bipartite_prod} suggest to downsample bipartite graphs by keeping all of the vertices in one subset of the bipartition, and \cite{shen_tree} suggests to downsample trees by keeping vertices at every other depth of the tree. Therefore, the largest eigenvector downsampling method can be seen as a generalization of those approaches.

\subsection{Connections with Graph Coloring and Spectral Clustering}
A graph $\G=\{\V,\E,\mathbf{W}\}$ is \emph{$k$-colorable} if there exists a partition of $\V$ into subsets $\V_1,\V_2,\ldots,\V_k$ such that if vertices $i,j\in \V$ are connected by an edge in $\E$, then $i$ and $j$ are in different subsets in the partition. The chromatic number $\chi$ of a graph $\G$ is the smallest $k$ such that $\G$ is $k$-colorable. Thus, the chromatic number of a graph is equal to 2 if and only if the graph is bipartite.

As we have seen with the examples in the previous section, when a graph is bipartite, it is easy to decide how to split it into two sets for downsampling.
 When the chromatic number of a graph is greater than two, however, we are interested in finding an \emph{approximate coloring} \cite{aspvall}; that is, a partition that has as few edges as possible that connect vertices in the same subset.\footnote{In other contexts, the term \emph{approximate coloring} is also used in reference to finding a proper $k$-coloring of a graph in polynomial time, such that $k$ is as close as possible to the chromatic number of the graph.} As noted by \cite{aspvall}, the approximate coloring problem is in some sense dual to the problem of spectral clustering (see, e.g. \cite{spectral_clustering} and references therein).

Aspvall and Gilbert \cite{aspvall} suggest to construct an approximate 2-coloring of an unweighted graph according to the polarity of the eigenvector associated with the most negative eigenvalue of the adjacency matrix. For  regular graphs, the eigenvector associated with the most negative eigenvalue of the adjacency matrix is the same as the largest graph Laplacian eigenvector, and so the method of \cite{aspvall} is equivalent  to  the largest Laplacian eigenvector method for that special case.

\subsection{Connections with Nodal Domain Theory}
A positive (negative) \emph{strong nodal domain} of $\mathbf{f}$ on $\G$ is a maximally connected subgraph such that ${f}(i)>0$ (${f}(i)<0$) for all vertices $i$ in the subgraph \cite[Chapter 3]{lap_eigen}. A positive (negative) \emph{weak nodal domain} of $\mathbf{f}$ on $\G$ is a maximally connected subgraph such that ${f}(i)\geq 0$ (${f}(i)\leq 0$) for all vertices $i$ in the subgraph, with ${f}(i)\neq 0$ for at least one vertex $i$ in the subgraph. 

A graph downsampling can be viewed as an assignment of positive and negative signs to vertices, with positive signs assigned to the vertices that we keep, and negative signs to the vertices that we eliminate. The goal of having few edges within either the removed set or the kept set is closely related to the problem of maximizing the number of nodal domains of the downsampling. This is because maximizing the number of nodal domains leads to nodal domains with fewer vertices, which results in fewer edges connecting vertices within the removed and kept sets.

Next, we briefly mention some general bounds on the 
number of nodal domains of eigenvectors of graph Laplacians: 
\begin{itemize}
\item[(ND1)] $\mathbf{u}_{\l}$ has at most $\l$ weak nodal domains and $\l+s-1$ strong nodal domains, where $s$ is the multiplicity of $\lambda_{\l}$ \cite{davies}, \cite[Theorem 3.1]{lap_eigen}.
\item[(ND2)] The largest eigenvector $\mathbf{u}_{\max}$ has $N$ strong and weak nodal domains if and only if $\G$ is bipartite. Moreover, if $\Hcal$ is an induced bipartite subgraph of $\G$ with the maximum number of vertices, then the number of vertices in $\Hcal$ is an upper bound on the number of strong nodal domains of any eigenvector of a generalized Laplacian of $\G$ \cite[Theorem 3.27]{lap_eigen}.
\item[(ND3)] If $\lambda_{\l}$ is simple and ${u}_{\l}(i) \neq 0,~\forall i \in \V$, the number of nodal domains of $\mathbf{u}_{\l}$ is greater than or equal to $\l-r$, where $r$ is the number of edges that need to be removed from the graph in order to turn it into a tree\footnote{Berkolaiko proves this theorem for Schr{\"o}dinger operators, which encompass generalized Laplacians of unweighted graphs.}\cite{berkolaiko}.
\end{itemize}

Note that while both the lower and upper bounds on the number of nodal domains of the eigenvectors of graph Laplacians are monotonic in the index of the eigenvalue, the actual number of nodal domains is not always monotonic in the index of the eigenvalue (see, e.g., \cite[Figure 1]{oren} for an example where they are not monotonic). Therefore, for arbitrary graphs, it is not guaranteed that the largest eigenvector of the graph Laplacian has more nodal domains than the other eigenvectors. For specific graphs such as bipartite graphs, however, this is guaranteed to be the case, as can be seen from the above bounds and Theorem \ref{Th:bipartite}. Finally, in the case of repeated graph Laplacian eigenvalues, \cite[Chapter 5]{lap_eigen} presents a hillclimbing algorithm to search for an associated eigenvector with a large number of nodal domains.

\subsection{Alternative Graph Downsampling Methods} \label{Se:alt_down}

We briefly mention some alternative graph downsampling/partitioning methods:
\begin{enumerate}
\item As mentioned above, \cite{aspvall} partitions the graph based on the polarity of the eigenvector associated with the most negative eigenvalue of the adjacency matrix. 
\item We can do a $k$-means clustering on $\mathbf{u}_{\max}$ with $k=2$ to separate the sets.
A closely related alternative is to set 
a non-zero threshold in \eqref{Eq:largest_eig_method}. A flexible choice of the threshold can ensure that $|\V_1|\approx \frac{N}{2}$.
\item In \cite{narang_icip},  Narang and Ortega downsample based on the solution to the weighted max-cut problem:
 \begin{align*}
\argmax_{\V_1} \Bigl\{\mbox{cut}(\V_1,\V_1^c)\Bigr\}=\argmax_{\V_1} \Biggl\{\sum_{i \in \V_1} \sum_{i^{\prime}\in \V_1^c} {W}_{ii^{\prime}}\Biggr\}.
 \end{align*}
 This problem is NP-hard, so approximations must be used for large graphs.
 \item Barnard and Simon \cite{barnard} choose $\V_1$ to be a maximal independent set (no edge connects two vertices in $\V_1$ and every vertex in $\V_1^c$ is connected to some vertex in $\V_1$). Maximal independent sets are not unique, but can be found via efficient greedy algorithms.
\item Ron et al. propose the SelectCoarseNodes algorithm \cite[Algorithm 2]{ron}, which leverages an algebraic distance measure to greedily select those nodes which have large future volume, roughly corresponding to those nodes strongly connected to vertices in the eliminated set.
\end{enumerate}
Additionally, there are a number of 
graph coarsening methods where the reduced set of vertices is not necessarily a subset of the original set of vertices. We discuss those further 
in Section \ref{Se:alternatives}.

\section{Graph Reduction} \label{Se:graph_reduction}
So we now have a way to downsample a graph to a subset of the vertices. However, in order to implement any type of multiscale transform that will require filtering operations defined on the subset of selected vertices, we still need to define a method to reduce the graph Laplacian on the original set of vertices to a new graph Laplacian on the subset of selected vertices. We refer to this process as \emph{graph reduction}. 
For the purpose of multiscale transforms, we would ideally like the graph reduction method to have some or all of the properties listed below. Note that the following represents one example of desiderata and is neither exhaustive nor definitive. 
\begin{itemize}
\item[(GR1)] The reduction is in fact a graph Laplacian (i.e., a symmetric matrix with row sums equal to zero and nonpositive off-diagonal elements).
\item[(GR2)] It preserves connectivity. That is, if the original graph is connected, then the reduced graph is also connected.
\item[(GR3)] The spectrum of the reduced graph Laplacian is representative of and contained in the spectrum of the original graph Laplacian.
\item[(GR4)] It preserves structural properties (e.g., if the original graph is a tree or bipartite, the reduced graph is accordingly a tree or bipartite).
\item[(GR5)] Edges in the original graph that connect vertices in the reduced vertex set are preserved in the reduced graph.
\item[(GR6)] It is invertible or partially invertible with some side information; i.e., from the reduced graph Laplacian and possibly some other stored information, we can recover the original graph Laplacian.
\item[(GR7)] It is computationally efficient to implement.
\item[(GR8)] It preserves graph sparsity. That is, the ratio of the number of edges over potential edges $\left(\frac{|{\cal E}|}{N(N-1)/2}\right)$ does not increase too much from the original graph to the reduced graph. In addition to preserving computational benefits associated with sparse matrix-vector multiplication, keeping this ratio low ensures that the reduced graph also captures local connectivity information and not just global information.
\item[(GR9)] There is some meaningful correspondence between the eigenvectors of the reduced graph Laplacian and the original graph Laplacian; for example, when restricted to the kept set of vertices, the lowest (smoothest) eigenvectors of the original Laplacian are similar to or the same as the eigenvectors of the reduced Laplacian.
\end{itemize} 
In the following, we discuss in detail one particular choice of graph reduction that satisfies some, but not all, of these properties. In Section \ref{Se:alternatives}, we briefly mention a few other graph reduction methods.
\subsection{Kron Reduction} \label{Se:kron_reduction}
\begin{figure}[tbh]
\centering
\centerline{\includegraphics[width=3.4in]{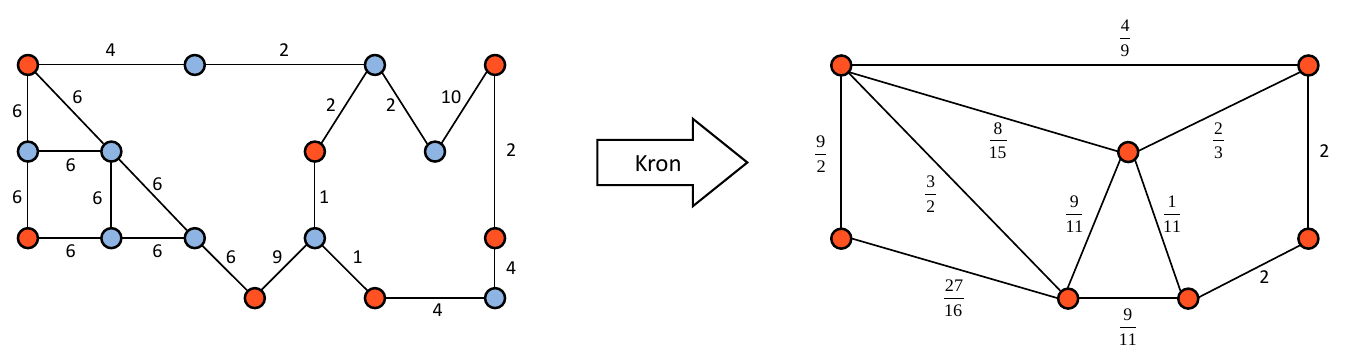}}
\caption {An example of Kron reduction. The vertices in $\V_1$ are in red, and those in $V_1^c$ are in blue.}
  \label{Fig:kron_example}
\end{figure}

The starting point for the graph reduction method we consider here is Kron reduction (see \cite{kron} and references therein). We start with a weighted graph $\G=\{\V,\E,\mathbf{W}\}$, its associated graph Laplacian $\L$, and a subset $\V_1 \subsetneq \V$ of vertices (the set selected by downsampling, for example) on which we want to form a reduced graph.\footnote{We assume $\card{\V_1}\geq 2$.}
The Kron reduction of $\L$ is the Schur complement \cite{schur} of $\L$ relative to $\L_{\V_1^c,\V_1^c}$; that is, the Kron reduction of $\L$ is given by
\begin{align}\label{Eq:kron_def}
\K\left(\L,\V_1\right):=\L_{\V_1,\V_1}-\L_{\V_1,\V_1^c}\L_{\V_1^c,\V_1^c}^{-1}\L_{\V_1^c,\V_1},
\end{align}
where $\L_{\Acal,\Bcal}$ denotes the $|\Acal| \times |\Bcal|$ submatrix consisting of all entries of $\L$ whose row index is in $\Acal$ and whose column index is in $\Bcal$.  
We can also uniquely associate with $\L^{Kron-reduced}=\K\left(\L,\V_1\right)$ 
a reduced weighted graph $\G^{Kron-reduced}=\left\{\V_1,\E^{Kron-reduced},\mathbf{W}^{Kron-reduced}\right\}$ by letting
\begin{align}\label{Eq:kron_weights}
{W}^{Kron-reduced}_{ij}=
\begin{cases}
- \L^{Kron-reduced}_{ij} & \text{if $i \neq j$,}
\\
0 &\text{o.w.}
\end{cases}
,
\end{align} and taking $\E^{Kron-reduced}$ to be the set of edges with non-zero weights in \eqref{Eq:kron_weights}.\footnote{In \eqref{Eq:kron_weights}, we still index the new weights by the same vertex indices $i$ and $j$. In practice, the vertices $i$ and $j$ will receive new ordinal indices between 1 and $N^{Kron-reduced}=|\V_1|$, the number of vertices in the reduced graph.} An example of Kron reduction is shown in Figure \ref{Fig:kron_example}.

\subsection{Properties of Kron Reduction}
For a complete survey of the topological, algebraic, and spectral properties of the Kron reduction process, see \cite{kron}. In the next theorem, we summarize the results of \cite{kron} that 
are of most interest with regards to multiscale transforms.
\begin{theorem}[D\"{o}rfler and Bullo, 2013]
Let $\L$ be a graph Laplacian of an undirected weighted graph $\G$ on the vertex set $\V$, $\V_1 \subsetneq \V$ be a subset of vertices with $\card{\V_1}\geq 2$, 
$\L^{Kron-reduced}=\K\left(\L,\V_1\right)$ be the Kron-reduced graph Laplacian defined in \eqref{Eq:kron_def}, and $\G^{Kron-reduced}$ be the associated Kron-reduced graph. Then
\begin{enumerate}
\item[(K1)] The Kron-reduced graph Laplacian in \eqref{Eq:kron_def} is well-defined
\item[(K2)] 
$\L^{Kron-reduced}$ is indeed a graph Laplacian
\item[(K3)] If the original graph $\G$ is connected $(\hbox{i.e., }\lambda_1(\L)>0)$, then the Kron-reduced graph $\G^{Kron-reduced}$ is also connected 
     $(\hbox{i.e., }\lambda_1(\L^{Kron-reduced})>0)$
\item[(K4)] If the original graph $\G$ is loopless, then the Kron-reduced graph $\G^{Kron-reduced}$ is also loopless\footnote{While \cite{kron} considers the more general case of \emph{loopy Laplacians}, the graphs we are interested in are all \emph{loopless}; that is, there are no self-loops connecting a vertex to itself, and therefore $\L_{nn}=\sum_{i\in \V, i \neq n}{W}_{ni}$.} 
\item[(K5)] Two vertices $i,j\in \V_1$ are connected by an edge if and only if there is a path between them in $\G$ whose vertices all belong to $\{i,j\} \cup \V_1^c$
\item [(K6)] Spectral interlacing: for all $\l \in \left\{0,1,\ldots,\card{\V_1}-1\right\}$,
\begin{align} \label{Eq:interlacing}
\lambda_{\l}(\L) \leq \lambda_{\l}(\L^{Kron-reduced})\leq \lambda_{\l+\card{\V}-\card{\V_1}}(\L).
\end{align}
In particular, \eqref{Eq:interlacing} and property (K4) above imply
\begin{align}\label{Eq:Kron_spectrum}
0=\lambda_{\min}(\L)&=\lambda_{\min}(\L^{Kron-reduced}) \nonumber \\
&\leq \lambda_{\max}(\L^{Kron-reduced}) \leq \lambda_{\max}(\L)
\end{align}
\item [(K7)] Monotonic increase of weights: for all $i,j \in \V_1$, ${W}^{Kron-reduced}_{ij} \geq {W}_{ij}$ 
\item [(K8)] Resistance distance \cite{klein} preservation: for all $i,j \in \V_1$, 
\begin{align*}
d_{R_{\G}}(i,j)&:=(\delta_i-\delta_j)^{\transpose}\L^{\dagger}(\delta_i-\delta_j)\\
&~=(\delta_i-\delta_j)^{\transpose}(\L^{Kron-reduced})^{\dagger}(\delta_i-\delta_j),
\end{align*}
where $\L^{\dagger}$ is the pseudoinverse of $\L$ and $\delta_i$ is the Kronecker delta (equal to 1 at vertex $i$ and 0 at all other vertices)
\end{enumerate}
\end{theorem}

\subsection{Special Cases}
As in Section \ref{Se:vs_special} for the vertex selection process, we analyze the effect of Kron reduction on some special classes of graphs in order to (i) provide further intuition behind the graph reduction process; (ii) show that Kron reduction is a generalization of previously suggested techniques for reducing graphs belonging to these special classes; and (iii) highlight both the strengths and weaknesses of this choice of graph reduction.
\subsubsection{Paths}
\begin{figure}[tb]
\centering
\begin{minipage}[b]{.45 \linewidth}
   \centering
   \centerline{\includegraphics[width=\linewidth]{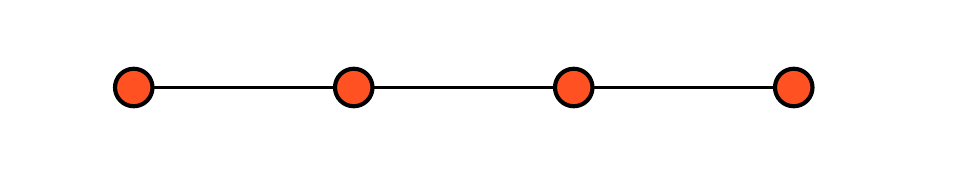}}
\centerline{\small{(a)}}
\end{minipage} 
\begin{minipage}[b]{.45 \linewidth}
   \centering
   \centerline{\includegraphics[width=\linewidth]{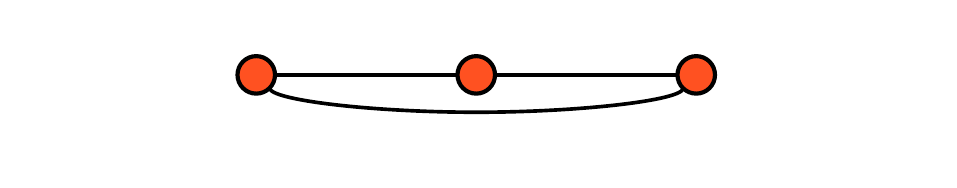}}
\centerline{\small{(b)}}
\end{minipage}\\
\vspace{0.4cm}
\begin{minipage}[b]{.45 \linewidth}
   \centering
   \centerline{\includegraphics[width=\linewidth]{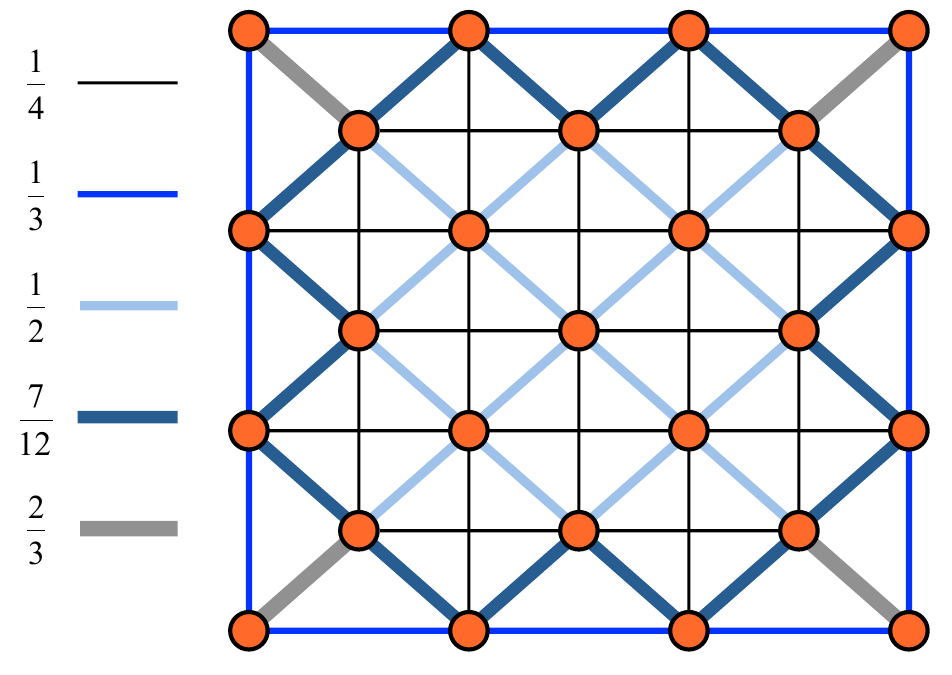}}
\centerline{\small{(c)}}
\end{minipage} 
\begin{minipage}[b]{.45 \linewidth}
   \centering
   \centerline{\includegraphics[width=\linewidth]{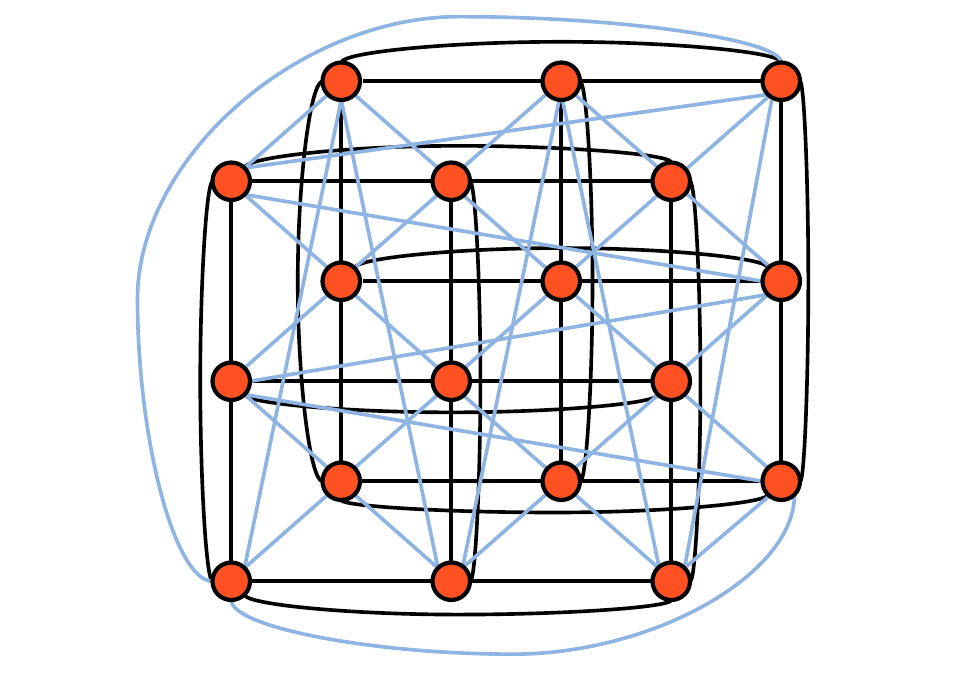}}
\centerline{\small{(d)}}
\end{minipage}\\
\vspace{0.4cm}
\begin{minipage}[b]{\linewidth}
   \centering
   \centerline{\includegraphics[width=.55\linewidth]{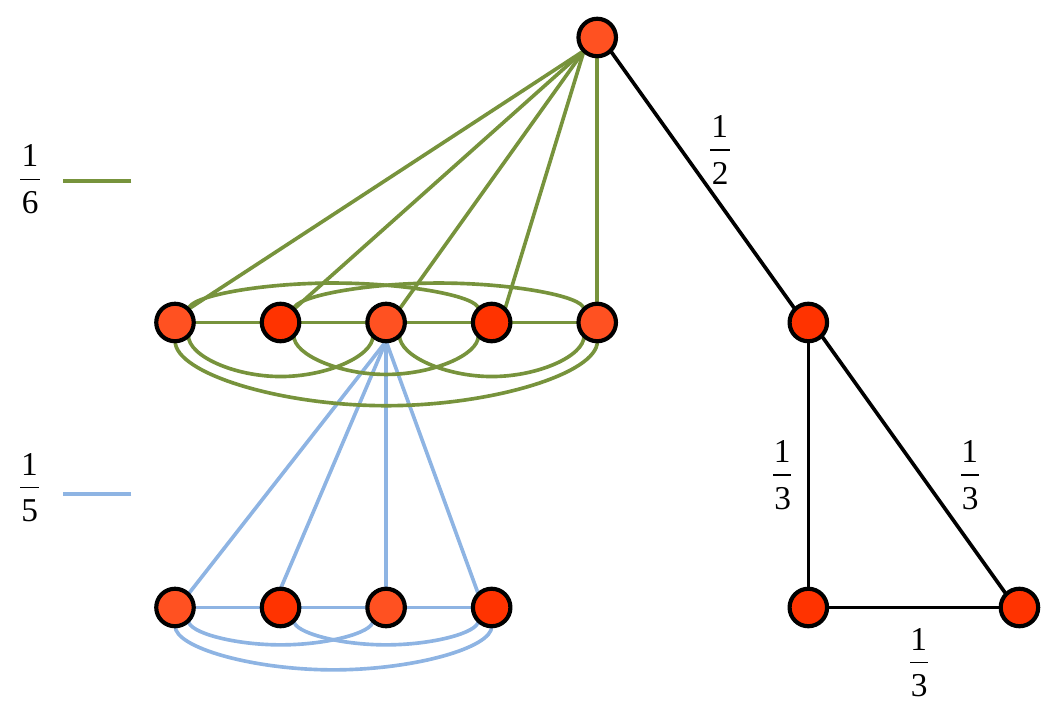}}
\centerline{\small{(e)}}
\end{minipage}
\caption {Kron reductions of the graphs of Figure \ref{Fig:special_graphs}, with the blue vertices eliminated. If the weights of all of the edges in the original path and ring graphs are equal to 1, then the weights of all of the edges in the Kron-reduced graphs in (a) and (b) are equal to $\frac{1}{2}$. For the Kron reduction of the finite grid graph with the ends connected (d), the weights of the blue edges are equal to $\frac{1}{2}$ and the weights of the black edges are equal to $\frac{1}{4}$. The weights of Kron reductions of the finite grid graph with the ends unconnected (c) and the tree graph (e) are shown by the color next to their respective Kron reductions.}
  \label{Fig:special_graphs_kron}
\end{figure}

The class of path graphs is closed under the sequential operations of largest eigenvector vertex selection and Kron reduction. That is, if we start with a path graph $\G_P$ with $N$ vertices, and select a subset $\V_1$ according the polarity of the largest eigenvector of the graph Laplacian of $\G_P$, then the graph associated with $\K\left(\G_P,\V_1\right)$ is also a path graph. If the weights of the edges of the original graph are all equal to $1$, the weights of the edges of the Kron-reduced graph are all equal to $\frac{1}{2}$. The Kron reduction of the graph shown in Figure \ref{Fig:special_graphs}(a) is shown in Figure \ref{Fig:special_graphs_kron}(a).

\subsubsection{Rings}
The Kron reduction of the ring graph shown in Figure \ref{Fig:special_graphs}(b) is shown in Figure \ref{Fig:special_graphs_kron}(b). It is a ring graph with half the number of vertices, and the weights of the edges are also halved in the Kron reduction process.
\subsubsection{Finite Grids}
The Kron reductions of the finite grid graphs shown in Figure \ref{Fig:special_graphs}(c,d) are shown in Figure \ref{Fig:special_graphs_kron}(c,d). If we rotate Figure \ref{Fig:special_graphs_kron}(c) clockwise by ninety degrees, we see that it is a subgraph of the 8-connected grid. We do not consider graphs with infinite vertices here, but the combined effect of the largest eigenvector vertex selection and Kron reduction operations is to map an infinite (or very large) 4-connected grid to an 8-connected grid.
\subsubsection{Trees} \label{Se:trees_kron}
The Kron reduction of the tree shown in Figure \ref{Fig:special_graphs}(e) is shown in Figure \ref{Fig:special_graphs_kron}(e).
Note that the Kron reduction of a tree is not a tree because siblings in the original graph are connected in the Kron-reduced graph. However, it does have a particular structure, which is a union of complete subgraphs. For every node $i$ in the reduced graph, and each one of that node's children $j$ in the original graph, there is a complete subgraph in the reduced graph that includes $i$ and all of the children of $j$ (i.e., the grandchildren of $i$ through $j$). If the weights of the edges in the original graph are equal to one, then the weights in each of the complete subgraphs are equal to one over the number of vertices in the complete subgraph. 

This class of graphs highlights two of the main weaknesses of the Kron reduction: (i) it does not always preserve regular structural properties of the graph; and (ii) it does not always preserve the sparsity of the graph.  We discuss 
a sparsity-enhancing modification in Section \ref{Se:sparsification}. 

\subsubsection{$k$-Regular Bipartite Graphs}
In \cite{narang_down}, Narang and Ortega consider connected and unweighted $k$-RBGs.\footnote{Although \cite{narang_down} considers unweighted $k$-RBGs, the following statements also apply to weighted $k$-RBGs if we extend the definition of the reduced adjacency matrix \eqref{Eq:red_adj} to weighted graphs.}
 They downsample by keeping one subset of the bipartition, and they construct a new graph on the downsampled vertices $\V_1$ by linking vertices in the reduced graph with an edge whose weight is equal to the number of their common neighbors in the original graph. If the vertices of the original graph are rearranged so that all the vertices in $\V_1$ have smaller indices than all the vertices in $\V_1^c$, the adjacency and Laplacian matrices of the original graph can be represented as:
\begin{align} \label{Eq:aSplit}
\mathbf{W}=\left[
\begin{array}{cc}
\mathbf{0} & \mathbf{W}_1 \\
\mathbf{W}_1^{\transpose} & \mathbf{0}
\end{array}
\right]
\hbox{ and }
\L=\left[
\begin{array}{cc}
k\mathbf{I}_{\frac{N}{2}} & -\mathbf{W}_1 \\
-\mathbf{W}_1^{\transpose} & k\mathbf{I}_{\frac{N}{2}}
\end{array}
\right].
\end{align}
Then for all $i,j \in \V_1$ (with $i \neq j$), the $(i,j)^{th}$ entry of the adjacency matrix of the reduced graph is given by
\begin{align} \label{Eq:red_adj}
{W}^{{kRBG-reduced}}_{ij}(\V_1)=({W}_1 {W}_1^{\transpose})_{ij}.
\end{align}
They also show that
\begin{align} \label{Eq:NO_red_L}
\L^{{kRBG-reduced}}(\V_1)=k^2 \mathbf{I}_{\frac{N}{2}} - \mathbf{W}_1 \mathbf{W}_1^{\transpose}.
\end{align}

Now we examine the Kron reduction of $k$-RBGs. The Kron-reduced Laplacian is given by:
\begin{align*}
\L^{Kron-reduced}(\V_1)&=\L_{\V_1,\V_1}-\L_{\V_1,\V_1^c}\L_{\V_1^c,\V_1^c}^{-1}\L_{\V_1^c,\V_1} \nonumber \\
&=k\mathbf{I}_{\frac{N}{2}}-(-\mathbf{W}_1)(k\mathbf{I}_{\frac{N}{2}})^{-1}(-\mathbf{W}_1^{\transpose}) \nonumber \\
&=k \mathbf{I}_{\frac{N}{2}} - \frac{1}{k}\mathbf{W}_1 \mathbf{W}_1^{\transpose},
\end{align*}
which is a constant factor $\frac{1}{k}$ times the reduced Laplacian \eqref{Eq:NO_red_L}
of \cite{narang_down}. So, up to a constant factor, the Kron reduction is a generalization of the graph reduction method presented in \cite{narang_down} for the special case of 
regular bipartite graphs.

\subsection{Graph Sparsification} \label{Se:sparsification}
As a consequence of property (K5), repeated Kron reduction often leads to denser and denser graphs. We have already seen this loss of sparsity in Section \ref{Se:trees_kron}, and 
this phenomenon is even more evident in larger, less regular graphs. In addition to computational drawbacks, the loss of sparsity can be important, because if the reduced graphs become too dense, they may not effectively capture the local connectivity information that is important for processing signals on the graph. Therefore, in many situations, it is advantageous to perform graph sparsification immediately after the Kron reduction as part of the overall graph reduction phase. 

\begin{figure}[tb]
\centering
\hfill
\begin{minipage}[b]{.25 \linewidth}
   \centering
   \centerline{\includegraphics[width=\linewidth]{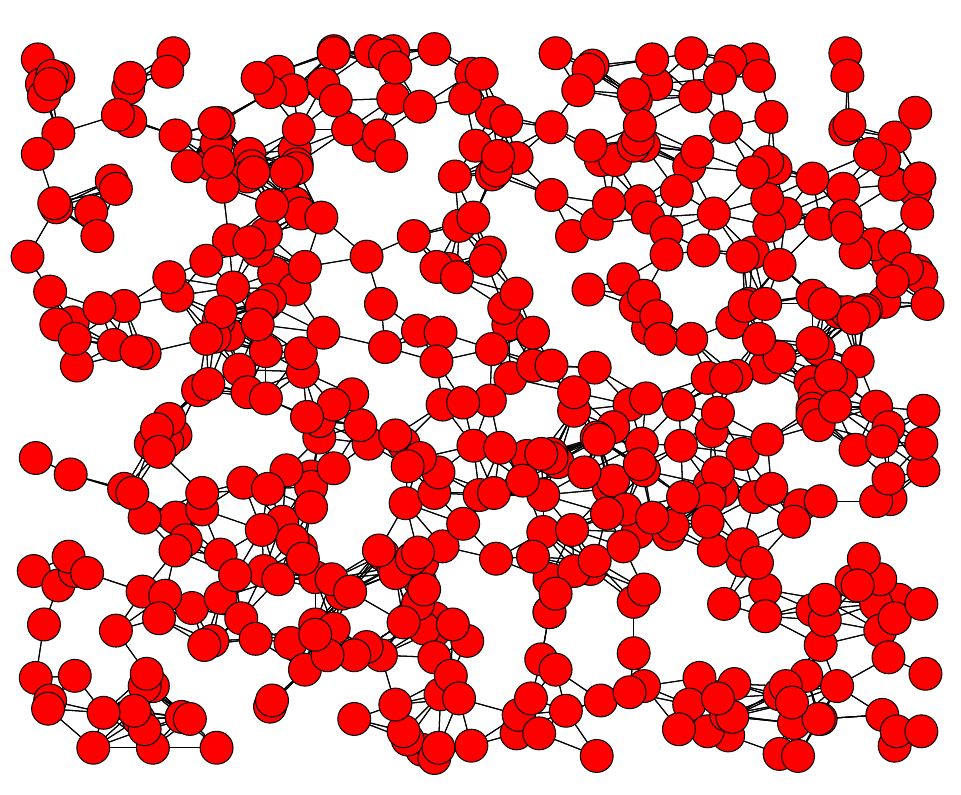}}
\centerline{\small{(a)}}
\end{minipage}
\hfill
\begin{minipage}[b]{.25 \linewidth}
   \centering
   \centerline{\includegraphics[width=\linewidth]{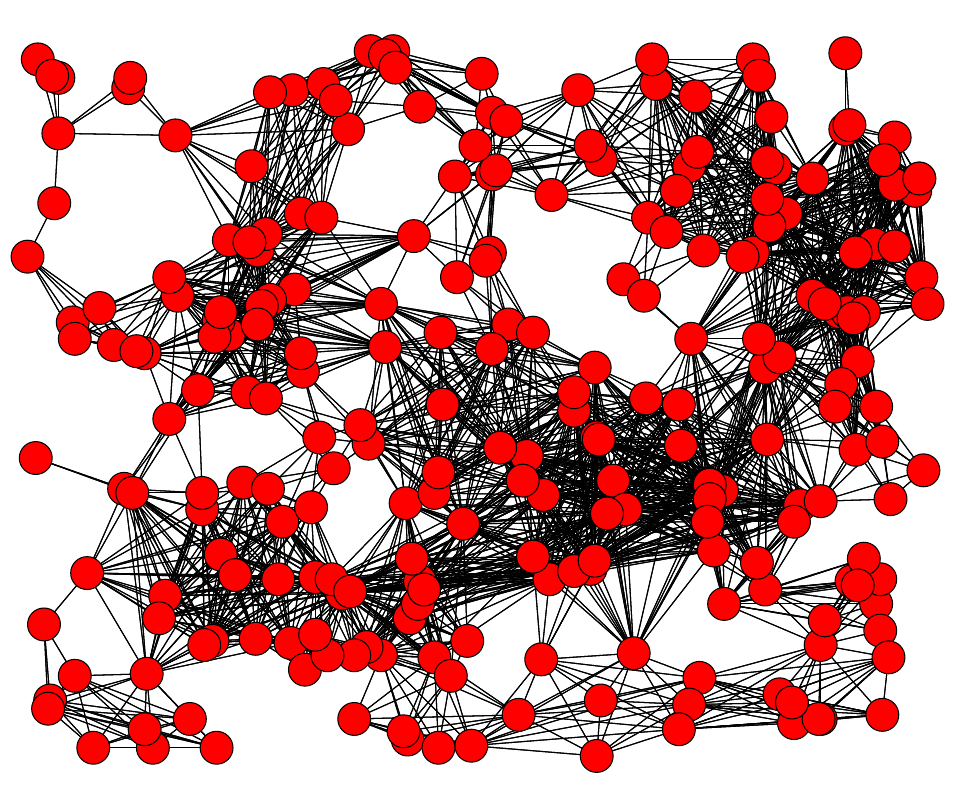}}
\centerline{\small{(b)}}
\end{minipage} 
\hfill
\begin{minipage}[b]{.25 \linewidth}
   \centering
   \centerline{\includegraphics[width=\linewidth]{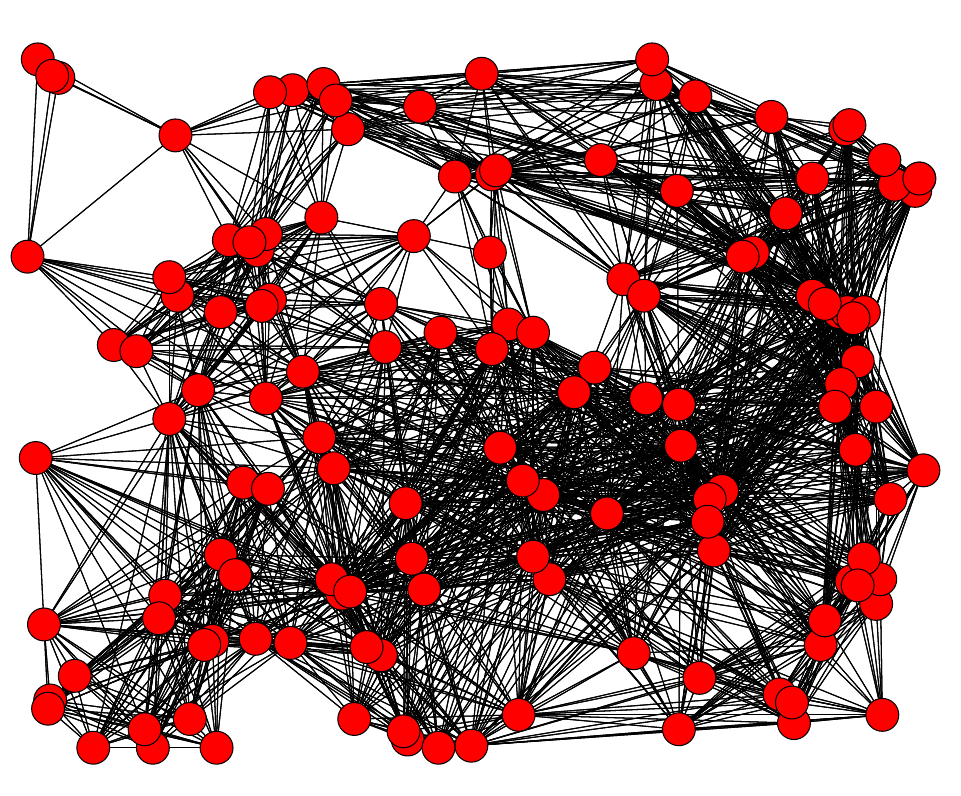}}
\centerline{\small{(c)}}
\end{minipage}
\hfill \\
\noindent
\hfill
\begin{minipage}[b]{.25 \linewidth}
   \centering
   \centerline{\includegraphics[width=\linewidth]{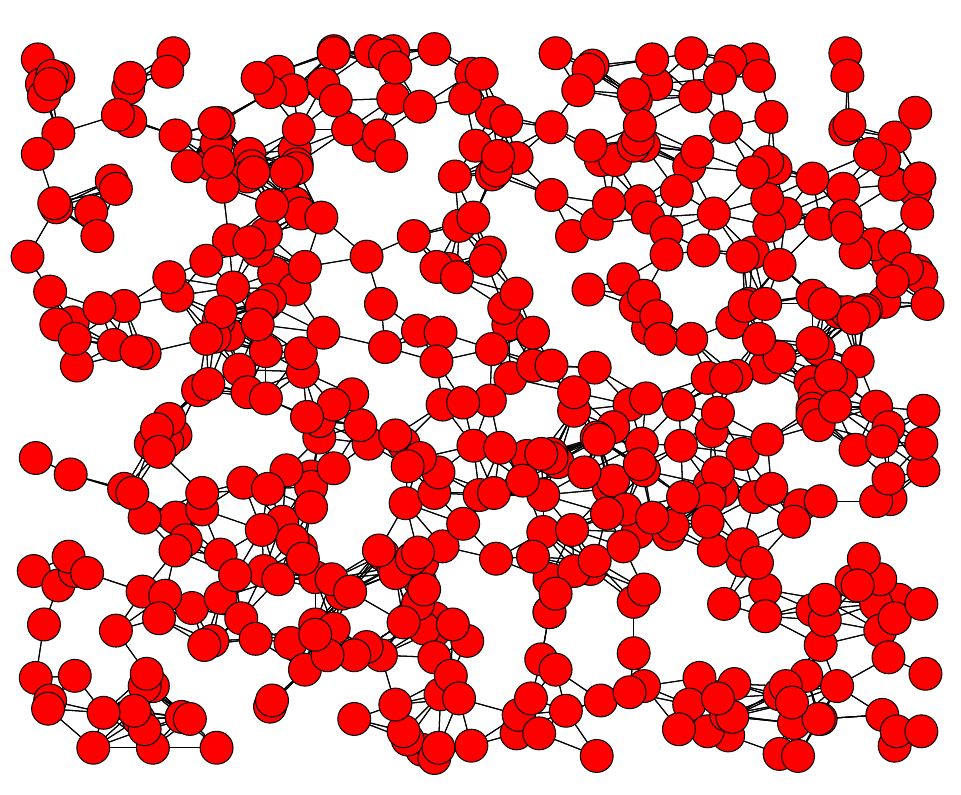}}
\centerline{\small{(d)}}
\end{minipage}
\hfill
\begin{minipage}[b]{.25 \linewidth}
   \centering
   \centerline{\includegraphics[width=\linewidth]{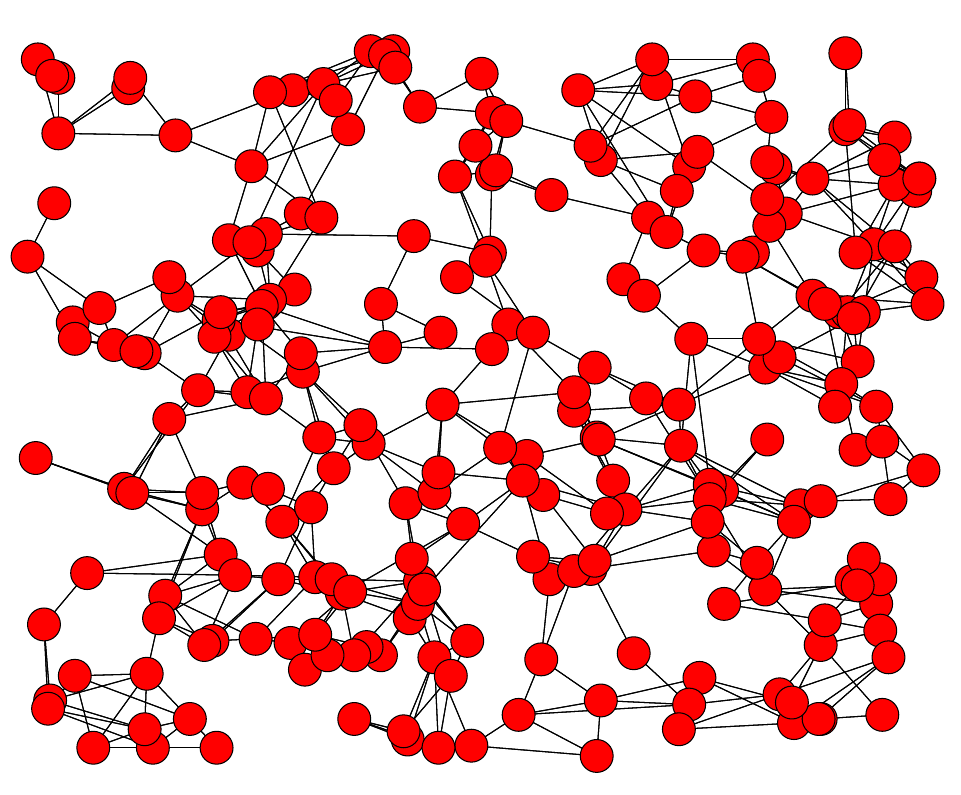}}
\centerline{\small{(e)}}
\end{minipage} 
\hfill
\begin{minipage}[b]{.25 \linewidth}
   \centering
   \centerline{\includegraphics[width=\linewidth]{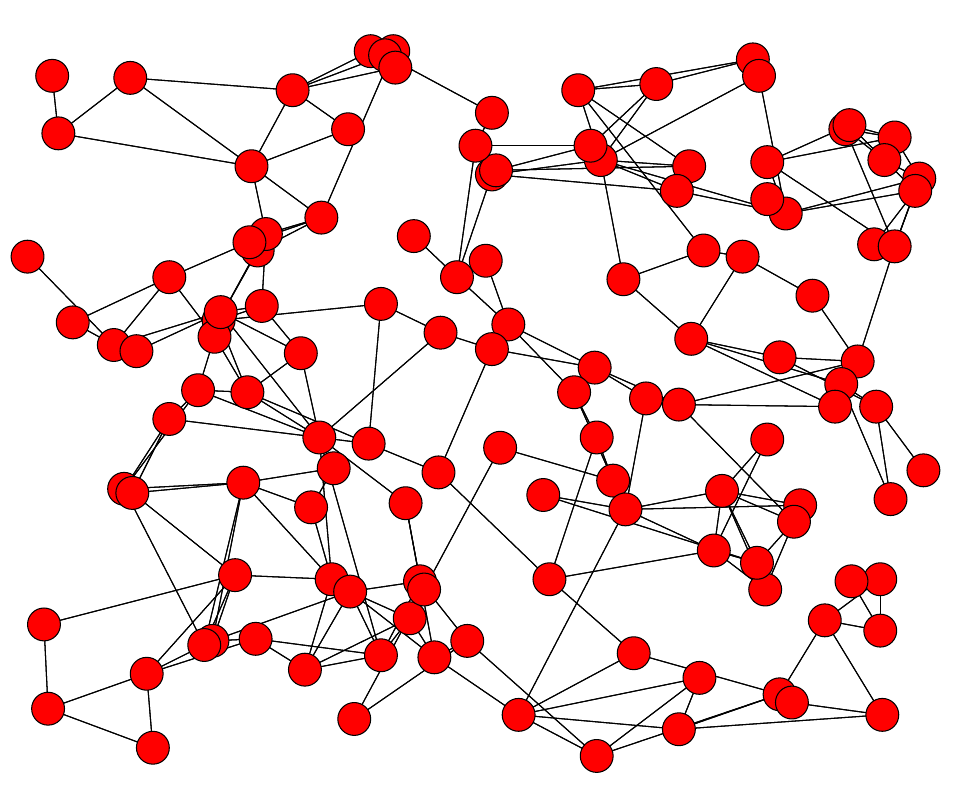}}
\centerline{\small{(f)}}
\end{minipage}
\hfill
\caption {Incorporation of a spectral sparsification step into the graph reduction. (a)-(c) Repeated largest eigenvector downsampling and Kron reduction of a sensor network graph. (d)-(f) The same process with the spectral sparsification of \cite{spielman_srivastava} used immediately after each Kron reduction.}
  \label{Fig:sparsification}
\end{figure}

There are numerous ways to perform graph sparsification. In this paper, we use a straightforward spectral sparsification algorithm of Spielman and Srivastava \cite{spielman_srivastava}, which is described in Algorithm \ref{Al:sparsification}.
This sparsification method pairs nicely with the Kron reduction, because \cite{spielman_srivastava} shows that for large graphs and an appropriate choice of the number of samples $Q$, the graph Laplacian spectrum and resistance distances between vertices are approximately preserved with high probability. In Figure \ref{Fig:sparsification}, we show an example of repeated downsampling followed by Kron reduction and spectral sparsification.

\begin{algorithm}[t] 
\caption{Spectral Sparsification \cite{spielman_srivastava}}
      Inputs: $\G=\{\V,\E,\mathbf{W}\}, Q$ \\
   Output:  $\mathbf{W}^{\prime}$ 
       \begin{algorithmic}[1]
   \STATE Initialize $\mathbf{W}^{\prime}=0$
   \FOR{$q=1,2,\ldots,Q$}
   \STATE Choose a random edge $e=(i,j)$ of $\E$ according to the probability distribution 
  \begin{align*}
  p_e=\frac{d_{R_{\G}}(i,j) W_{ij}}{\sum\limits_{e=(m,n) \in \E}d_{R_{\G}}(m,n) W_{mn}}
  \end{align*}
   \STATE ${W}^{\prime}_{ij}={W}^{\prime}_{ij}+\frac{{W}_{ij}}{Q p_e}$
   \ENDFOR
   \end{algorithmic} \label{Al:sparsification}
   \end{algorithm}

\subsection{Alternative Graph Reduction Methods} \label{Se:alternatives}
First, we mention some alternative graph reduction methods:
\begin{enumerate}
\item
In \cite{narang_icip}, Narang and Ortega define a reduced graph via the weighted adjacency matrix by taking $\mathbf{W}^{(j+1)}=\left([\mathbf{W}^{(j)}]^2\right)_{\V_1,\V_1}$. Here, 
$[\mathbf{W}^{(j)}]^2$
represents the 2-hop adjacency matrix of the original graph. The reduced Laplacian can then be defined as $\mathbf{\L}^{(j+1)}=\mathbf{D}^{(j+1)}-\mathbf{W}^{(j+1)}$, where $\mathbf{D}^{(j+1)}$ is computed from $\mathbf{W}^{(j+1)}$. However, there are a number of undesirable properties of this reduction method. First, and perhaps foremost, the reduction method does not always preserve connectivity. Second, self-loops are introduced at every vertex in the reduced graph. Third, vertices in the selected subset that are connected by an edge in the original graph may not share an edge in the reduced graph. Fourth, the spectrum of the reduced graph Laplacian is not necessarily contained in the spectrum of the original graph Laplacian.
\item Ron et al. \cite{ron} assign a fraction $P_{ij}$ of each vertex $i$ in the original graph  to each vertex 
$j \in \V_1$ in the reduced graph that is close to $i$ in the original graph (in terms of an algebraic distance). The assignment satisfies $\sum_{j \in \V_1} P_{ij}=1$ for all $i\in \V$ and $P_{jj}=1$ for all $j \in \V_1$, and for each eliminated vertex $i \in \V_1^c$, an upper limit is placed on the number of vertices $j \in \V_1$ that have $P_{ij}>0$. Then for all $j,j^{\prime} \in \V_1$,
\begin{align*}
W_{jj^{\prime}}^{reduced}:=\sum_{m,n \in \V, m \neq n} P_{mj} P_{nj^{\prime}} W_{mn}. 
\end{align*}
\end{enumerate}
Next, we mention some graph coarsening (also called coarse-graining) methods that combine graph downsampling and reduction into a single operation by forming aggregate nodes at each resolution, rather than keeping a strict subset of original vertices. The basic approach of these methods is to partition the original set of vertices into clusters, represent each cluster of vertices in the original graph with a single vertex in the reduced graph, and then use the original graph to form edges and weights that connect the representative vertices in the reduced graph. Some examples include:
\begin{enumerate}
\item  Lafon and Lee \cite{lafon_coarse} cluster based on diffusion distances and form new edge weights based on random walk transition probabilities.   
\item The Lean Algebraic Multigrid (LAMG) method of \cite{livne} builds each coarser Laplacian by selecting seed nodes, assigning each non-seed node to be aggregated with exactly one seed node, and setting the weights between two new seed nodes $j$ and $j^{\prime}$ to be 
$$ W_{jj^{\prime}}^{reduced}:=\sum_{i \in \V_j} \sum_{i^{\prime} \in \V_{j^{\prime}}} W_{ii^{\prime}},$$
where $\V_j$ and $\V_{j^{\prime}}$ are the sets of vertices that have been aggregated with seeds $j$ and $j^{\prime}$, respectively. The seed assignments are based on an affinity measure that approximates short time diffusion distances, rather than the algebraic distances used in \cite{ron}.

\item Multilevel clustering algorithms such as those presented in \cite{METIS,dhillon} often use greedy coarsening algorithms such as heavy edge matching or max-cut coarsening (see \cite[Section 3]{METIS} and \cite[Section 5.1]{dhillon} for details).
\end{enumerate}
For more thorough reviews of the graph partitioning and coarsening literature, see, e.g., \cite{teng, ron,gp_archive}. 

To summarize, given a graph $\G$ and a desired number of resolution levels, in order to generate a multiresolution of graphs, we only need to choose a downsampling operator and a graph reduction method. In the remainder of the paper, we use the largest eigenvector downsampling operator and the Kron reduction with the extra sparsification step
to generate graph multiresolutions such as the three level one shown in Figure \ref{Fig:sparsification}(d)-(f). Note that graph multiresolutions generated in this fashion are completely independent of any signals residing on the graph. 
  
\section{Filtering and Interpolation of Graph Signals}\label{Se:filtering}
Equipped with a graph multiresolution, we now proceed to analyze signals residing on the finest graph in the multiresolution. The two key graph signal processing components we use in the proposed transform are a generalized filtering operator and an interpolation operator for signals on graphs. 
Filters are commonly used in classical signal processing analysis to separate a signal into different frequency bands. In this section, we review how to extend the notion of filtering to graph signals. We focus on graph spectral filtering, which leverages the eigenvalues and eigenvectors of Laplacian operators from spectral graph theory \cite{chung} to capture the geometric structure of the underlying graph data domain. We then discuss different methods to interpolate from a signal residing on a coarser graph to a signal residing on a finer graph whose vertices are a superset of those in the coarser graph. 

\subsection{Graph Spectral Filtering} \label{Se:graph_spectral_filtering}
A \emph{graph signal} is a function $f: \V \rightarrow \Rbb$ that associates a real value to each vertex of the graph.
Equivalently, we can view a graph signal as a vector $f \in \Rbb^N$.  

In frequency filtering, we represent signals as linear combinations of a set of 
signals and amplify or attenuate the contributions of 
different components.
In classical signal processing, the set of 
component signals are usually the complex exponentials, which carry a notion of frequency and give rise to the Fourier transform. In graph signal processing, it is most common to choose the graph Fourier expansion basis to be the eigenvectors of the combinatorial or normalized graph Laplacian operators. This is because the spectra of these graph Laplacians also carry a notion of frequency (see, e.g., \cite[Figure 3]{shuman_SPM}), and their eigenvectors are the graph analogs to the complex exponentials, which are the eigenfunctions of the classical Laplacian operator.

More precisely, the graph Fourier transform with the combinatorial graph Laplacian eigenvectors as a basis is
\begin{align}\label{Eq:graph_FT}
\hat{f}(\lambda_{\l}) := \langle {\mathbf{f},\mathbf{u}_{\l}}\rangle = \sum_{i=1}^N f(i) u_{\l}^*(i), 
\end{align}
and a \emph{graph spectral filter}, which we also refer to as a kernel, is a real-valued mapping $\hat{h}(\cdot)$ on the spectrum of graph Laplacian eigenvalues. Just as in classical signal processing, the effect of the filter is multiplication in the Fourier domain: 
\begin{align} \label{Eq:filtering1}
\hat{f}_{out}(\lambda_{\l}) = \hat{f}_{in}(\lambda_{\l})\hat{h}(\lambda_{\l}),
\end{align}
or, equivalently, taking an inverse graph Fourier transform, 
\begin{align} \label{Eq:filtering2}
f_{out}(i) = \sum_{\l=0}^{N-1} \hat{f}_{in}(\lambda_{\l})\hat{h}(\lambda_{\l}) u_{\l}(i).
\end{align}
We can also write the filter in matrix form as 
$\mathbf{f}_{out}=\mathbf{H}\mathbf{f}_{in}$, where $\mathbf{H}$ is 
a matrix function \cite{higham}
\begin{align}\label{Eq:mat_form}
\mathbf{H}=\hat{h}(\L)=\mathbf{U}[\hat{h}(\mathbf{\Lambda})]\mathbf{U}^{*},
\end{align} 
where $\hat{h}(\mathbf{\Lambda})$ is a diagonal matrix with the elements of the diagonal equal to $\{\hat{h}(\lambda_{\l})\}_{\l=0,1,\ldots,N-1}$. We can also use the normalized graph Laplacian eigenvectors as the graph Fourier basis, and simply replace $\L$, $\lambda_{\l}$, and $\mathbf{u}_{\l}$ by $\tilde{\L}$, $\tilde{\lambda}_{\l}$, and $\tilde{\mathbf{u}}_{\l}$ in \eqref{Eq:graph_FT}-\eqref{Eq:mat_form}. A discussion of the benefits and drawbacks of each of these choices for the graph Fourier basis is included in \cite{shuman_SPM}.

\subsection{Alternative Filtering Methods for Graph Signals}
We briefly mention two alternative graph filtering methods:
\begin{enumerate}
\item We can filter a graph signal directly in the vertex domain by writing the output at a given vertex $i$ as a linear combination of the input signal components in a neighborhood of $i$. Graph spectral filtering with an order $K$ polynomial kernel can be viewed as filtering in the vertex domain with the component of the output at vertex $i$ written as a linear combination of the input signal components in a $K$-hop neighborhood of $i$ (see \cite{shuman_SPM} for more details).
\item Other choices of filtering bases can be used in place of $\L$ in \eqref{Eq:mat_form}. For example, in \cite{sandryhaila_2013,sandryhaila_2014}, Sandryhaila and Moura examine filters that are polynomial functions of the adjacency matrix. 
\end{enumerate}

\subsection{Interpolation on Graphs} \label{Se:interpolation}
Given the values of a graph signal on a subset, ${\V_1}$, of the vertices, the interpolation problem is to infer the values of the signal on the remaining vertices, ${\V_1}^c=\V \backslash {\V_1}$. It is usually assumed that the signal to be interpolated is smooth. This assumption is always valid in our construction, because the signals we wish to interpolate have been smoothed via a lowpass graph spectral filter.

\begin{figure}[t]
\centering
\hfill
\begin{minipage}[b]{.31 \linewidth}
   \centering
   \centerline{\includegraphics[width=\linewidth]{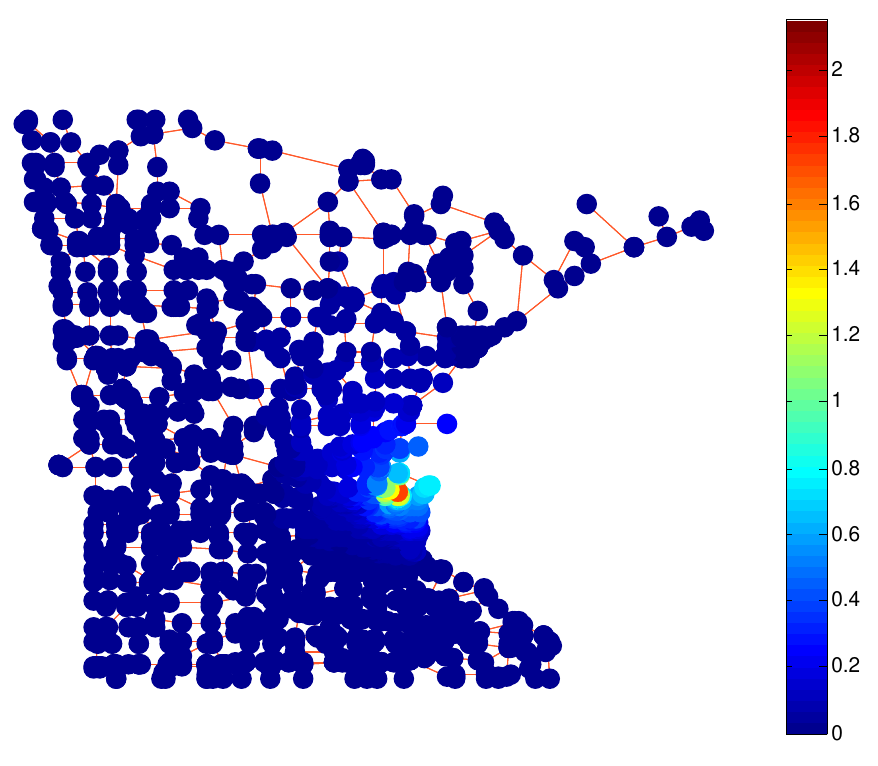}}
\end{minipage}
\hfill
\begin{minipage}[b]{.31 \linewidth}
   \centering
   \centerline{\includegraphics[width=\linewidth]{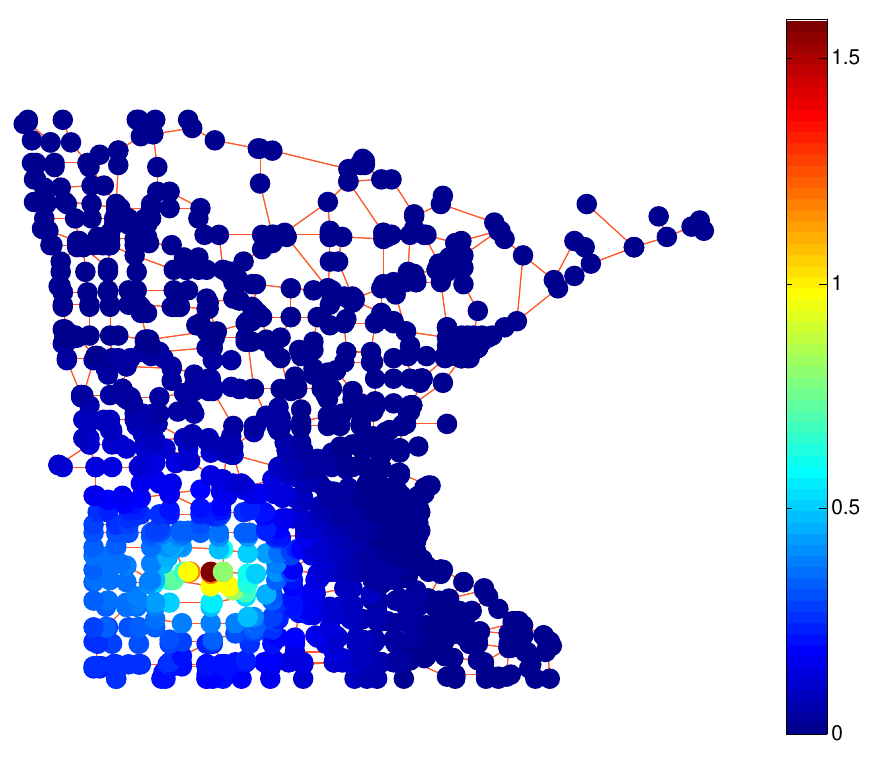}}
\end{minipage} 
\hfill
\begin{minipage}[b]{.31 \linewidth}
   \centering
   \centerline{\includegraphics[width=\linewidth]{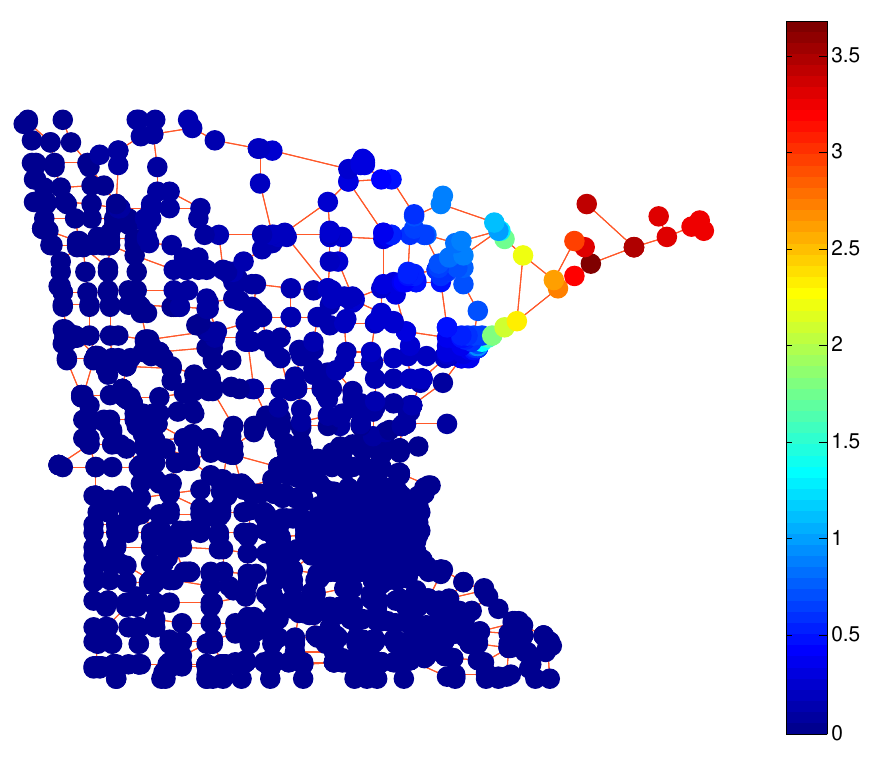}}
\end{minipage}
\hfill
\caption {Three localized Green's functions, centered at different vertices on the Minnesota graph.}
  \label{Fig:greens}
\end{figure}

We use the interpolation model of Pesenson \cite{pesenson_splines}, which is to interpolate the missing values, $\mathbf{f}_{\V_1^c}$, of the signal $\mathbf{f}$ by writing the interpolant as a linear combination of the Green's functions of the regularized graph Laplacian, $\bar{\L}:=\L+\epsilon \mathbf{I}$, that are centered at the vertices in $\V_1$.\footnote{Pesenson's model and subsequent analysis in \cite{pesenson_splines} is for any power $\bar{\L}^t$ of the regularized Laplacian, but we stick to $t=1$ throughout for simplicity.} That is, the interpolation is given by 
\begin{align}\label{Eq:interp_model}
\mathbf{f}_\textrm{interp}= \sum_{j \in \mathcal{V}_r} \alpha[j] {\boldsymbol{\varphi}}_j=\mathbf{\Phi}_{\mathcal{V}_1} \mathbf{\alpha},
\end{align}
where ${\boldsymbol{\varphi}}_j$ is a Green's kernel $\hat{g}(\lambda_{\l})=\frac{1}{\lambda_{\l}+\epsilon}$ translated to center vertex $j$ (see \cite{sgwt,shuman_SPM} for more on generalized translation on graphs):
\begin{align*}
{\boldsymbol{\varphi}}_j=T_j g = \hat{g}(\L)\delta_j =\mathbf{U}[\hat{g}(\mathbf{\Lambda})]\mathbf{U}^* \delta_j = \sum_{\l=0}^{|\V|-1} \frac{1}{\lambda_{\l}+\epsilon} u_{\l}^*(j) \mathbf{u}_{\l},
\end{align*}
and $\mathbf{\Phi}_{\mathcal{V}_1}$ is a $|\V| \times |\V_1|$ matrix with columns $\{{\boldsymbol{\varphi}}_j\}_{j \in \V_1}$. We display three such translated Green's functions in Figure \ref{Fig:greens}. 
Note that these are localized functions satisfying the property
\begin{align*}
\bar{\L}{\boldsymbol{\varphi}}_j = \Bigl(\mathbf{U}[\mathbf{\Lambda} +\epsilon \mathbf{I}]\mathbf{U}^*\Bigr)\Bigl( \mathbf{U}[\hat{g}(\mathbf{\Lambda})]\mathbf{U}^* \delta_j \Bigr)= \bar{\L}\bar{\L}^{-1}\delta_j = \delta_j.
\end{align*}
The parameter $\epsilon$ controls the spread of the translated Green's functions around their center vertices, with a smaller value of $\epsilon$ leading to larger spreads.

To perform the interpolation, we find the coefficients $\mathbf{\alpha}$ such that the linear combination of Green's functions restricted to the subset $\V_1$ of vertices yields the signal values on this subset:
\begin{align}\label{Eq:interp_eq}
\mathbf{f}_{\V_1}=\mathbf{\Phi}_{\V_1,\V_1} \mathbf{\alpha_*},
\end{align}
where $\mathbf{f}_{\V_1}$ is a vector containing the values of the signal on the vertices in $\V_1$, and $\mathbf{\Phi}_{\V_1,\V_1}$ is the submatrix that contains only the rows of $\mathbf{\Phi}_{\V_1}$ corresponding to the vertices in $\V_1$.
To compute $\mathbf{\alpha}_*=\mathbf{\Phi}_{\V_1,\V_1}^{-1} \mathbf{f}_{\V_1}$, we note that 
\begin{align*}
\mathbf{\Phi}_{\V_1,\V_1} &= \left(\bar{\L}^{-1}\right)_{\V_1,\V_1} \\ &=\left[\bar{\L}_{\V_1,\V_1}-\bar{\L}_{\V_1,\V_1^c}\left(\bar{\L}_{\V_1^c,\V_1^c}\right)^{-1}\bar{\L}_{\V_1^c,\V_1}\right]^{-1},
\end{align*}
where the last equality follows from the Matrix Inversion Lemma. Thus, the cost to compute 
\begin{align}\label{Eq:interp_kron}
\mathbf{\alpha}_*=\left[\bar{\L}_{\V_1,\V_1}-\bar{\L}_{\V_1,\V_1^c}\left(\bar{\L}_{\V_1^c,\V_1^c}\right)^{-1}\bar{\L}_{\V_1^c,\V_1}\right]\mathbf{f}_{\V_1}
\end{align}
is dominated by the solution of a sparse diagonally dominant set of $|\V|-|\V_1|$ linear equations.
Finally, we can substitute $\mathbf{\alpha}_*$ back into \eqref{Eq:interp_model} to find the interpolated signal, whose values on $\V_1$ of course correspond to the given values $\mathbf{f}_{\V_1}$. This last calculation can be approximated efficiently by upsampling $\mathbf{\alpha}_*$ to $\V$, and filtering it with the Green's kernel $\hat{g}(\cdot)$ using the polynomial approximation algorithm discussed in Section \ref{Se:chebyshev}. Pesenson \cite{pesenson_splines} refers to the interpolated signal $\mathbf{f}_\textrm{interp}$ as a \emph{variational spline}.

As a brief aside, we note that the matrix in \eqref{Eq:interp_kron} is exactly the Kron reduction $\K(\bar{\L},\V_1)$ of the regularized graph Laplacian on the set of vertices $\V_1$ where the signal values are known.

As with downsampling and graph reduction, there are a number of different methods to interpolate smooth signals on graphs, and we briefly mention a few alternatives here:
\begin{enumerate}
\item Graph-based semi-supervised learning methods, such as Tikhonov regularization (see, e.g., \cite{belkin_matveeva}-\nocite{smola, reg_discrete, harmonic,zhou_bousquet}\cite{chapelle}). Like many of those methods, the spline interpolation method we use can be viewed as solving an optimization problem (see \cite{pesenson_splines}). One main difference is that in our multiscale transform, we will usually have approximately half of the signal values available, whereas the number of signal values available in the semi-supervised learning literature is usually assumed to be significantly smaller.
\item Grady and Schwartz's anisotropic interpolation \cite{grady_interpolation} solves the system of equations
 $\L_{\V_1^c,\V_1^c}\mathbf{f}_{\V_1^c}=-\L_{\V_1^c,\V_1}\mathbf{f}_{\V_1}$ in order to minimize $\mathbf{f}_\textrm{interp}^{\transpose} \L \mathbf{f}_\textrm{interp}$ subject to ${f}_\textrm{interp}(i)=f(i)$ for all $i \in \V$.
\item Rather than using the Green's functions as the interpolating functions in \eqref{Eq:interp_model}, Narang et al. \cite{narang_interp1,narang_interp2} use the eigenvectors of the normalized graph Laplacian associated with the lowest eigenvalues. 
\end{enumerate}

\section{A Pyramid Transform for Signals on Graphs} \label{Se:pyramid}
We are finally ready to combine the downsampling, 
graph reduction, filtering, and interpolation operations from the previous three sections to define a 
multiscale pyramid transform for signals on graphs. After reviewing the classical Laplacian pyramid, we summarize our extension, first for a single level and then for the whole pyramid.

\subsection{The Classical Laplacian Pyramid}
\begin{figure}[tb]
\centering
\begin{minipage}[b]{\linewidth}
   \centering
   \centerline{\includegraphics[width=.8\linewidth]{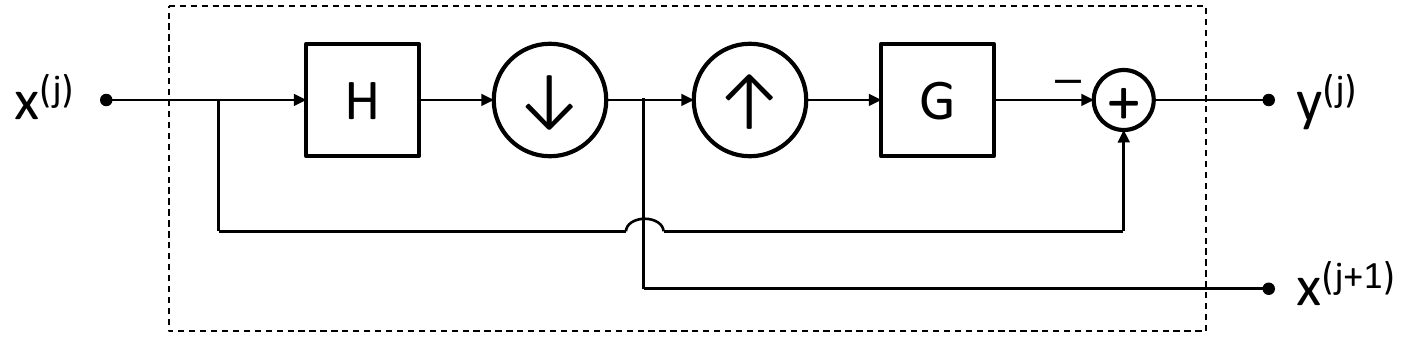}}
\centerline{\small{(a)}}
\end{minipage} \\
\vspace{0.4cm}
\begin{minipage}[b]{\linewidth}
   \centering
   \centerline{\includegraphics[width=.8\linewidth]{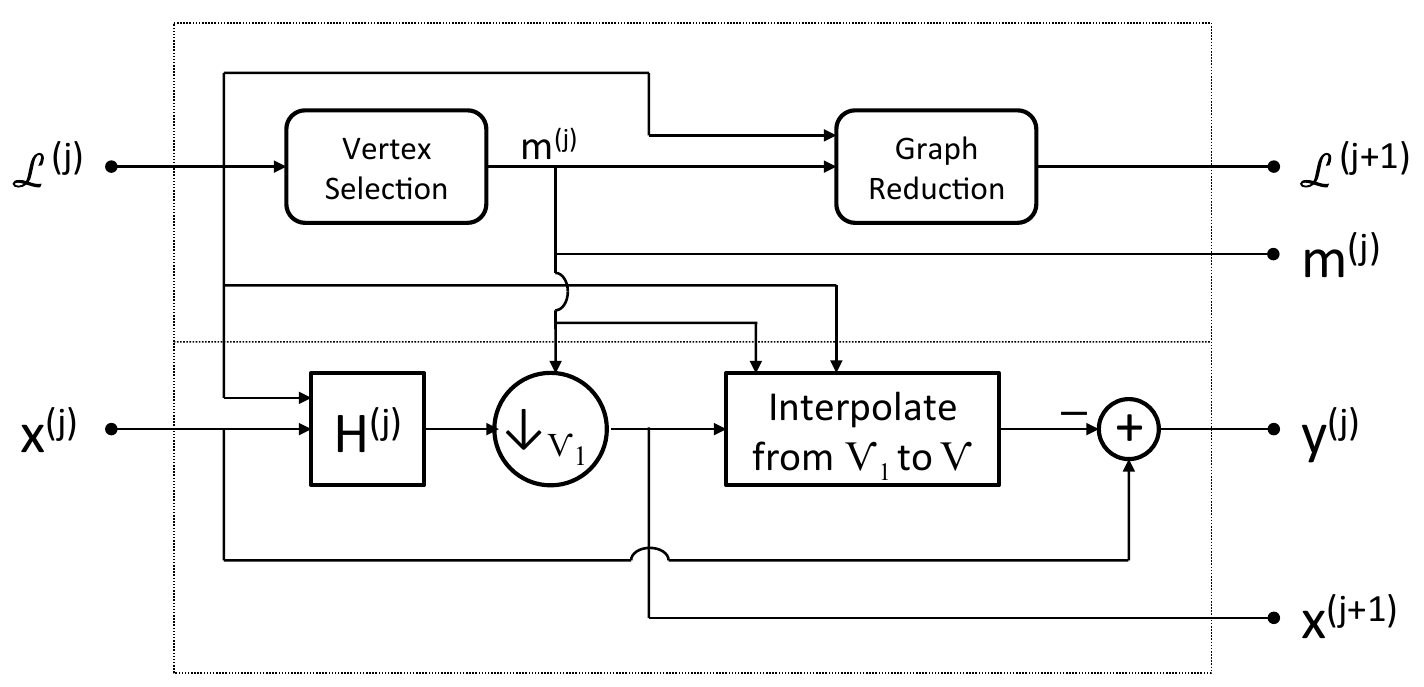}}
\centerline{\small{(b)}}
\end{minipage}\\
\vspace{0.4cm}
\begin{minipage}[b]{\linewidth}
   \centering
   \centerline{\includegraphics[width=.8\linewidth]{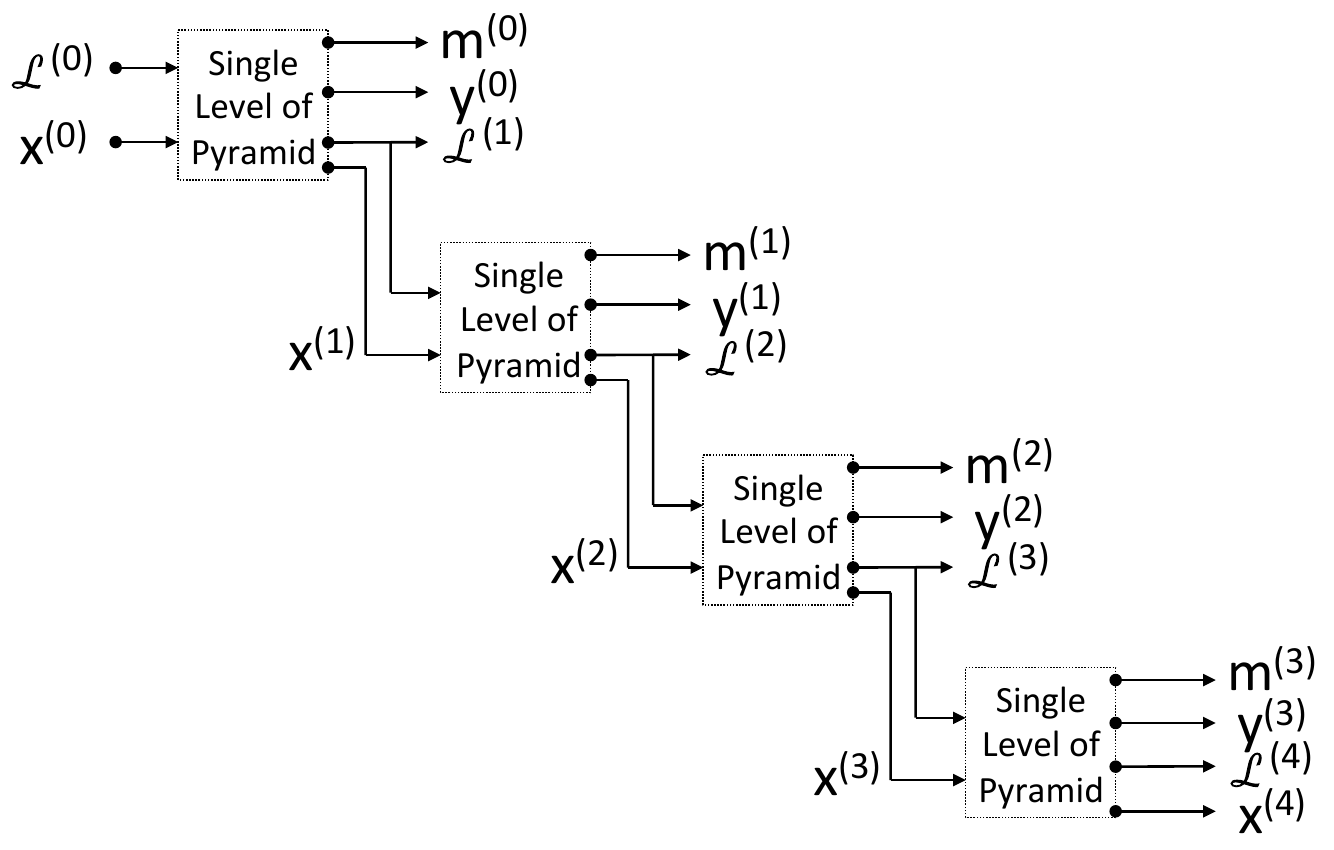}}
\centerline{\small{(c)}}
\end{minipage}
\caption {(a) A single level of the classical Laplacian pyramid of Burt and Adelson \cite{burt_adelson}. (b) A single level of the pyramid scheme proposed for signals defined on graphs. The top half consisting of the graph downsampling and reduction is specific to the graph (but the same for every signal on that graph), while the bottom half is specific to each graph signal. (c) A multilevel pyramid for signals defined on graphs.}
  \label{Fig:lp}
\end{figure}

In \cite{burt_adelson}, Burt and Adelson introduce the Laplacian pyramid. Originally designed with image coding in mind, the Laplacian pyramid is a multiresolution transform that is applicable to any regularly-sampled signal in time or space. A schematic representation of a single level of the Laplacian pyramid is shown in Figure \ref{Fig:lp}(a). At each level of the pyramid, an input signal $\mathbf{x}$ is lowpass filtered ($\mathbf{H}$) and then downsampled to form a coarse approximation of the original signal. A prediction of the input signal is then formed by upsampling and lowpass filtering ($\mathbf{G}$) the coarse approximation. The prediction error $\mathbf{y}$, i.e., the difference between the input signal and the prediction based on the coarse approximation, is stored for reconstruction. This process is iterated, with the coarse approximation that is output at the previous level acting as the input to the next level. The sequence $\left\{\mathbf{x}^{(i)}\right\}_{i=1,2,\ldots}$ represents a series of coarse approximations of the original signal $\mathbf{x}^{(0)}$ at decreasing resolutions. If, for example, the downsampling is by a factor of 2 and the original signal $\mathbf{x}^{(0)} \in \Rbb^N$, then $\mathbf{x}^{(i)} \in \Rbb^{2^{-i}N}$.

The lowest resolution approximation, $\mathbf{x}^{(J)}$, of the original signal and the sequence of prediction errors, $\left\{\mathbf{y}^{(i)}\right\}_{i=0,1,\ldots,J-1}$, are stored and/or transmitted for reconstruction. Therefore, the Laplacian pyramid is an overcomplete transform, as a $J$-level pyramid downsampled by a factor of $\kappa$ at each level maps an input of dimension $N$ into $N\left(\frac{1-\frac{1}{\kappa}^{J+1}}{1-\frac{1}{\kappa}}\right)$ transform coefficients. On the other hand, the Laplacian pyramid has a number of desirable properties. First, perfect reconstruction of the original signal is possible for any choice of filtering operations $\mathbf{H}$ and $\mathbf{G}$. It is easy to see that upsampling $\mathbf{x}^{(j+1)}$, filtering it by $\mathbf{G}$, and summing with the prediction error $\mathbf{y}^{(j)}$ yields the finer approximation $\mathbf{x}^{(j)}$.
Second, the prediction errors usually have less entropy than the original signal of the same dimension, enabling further compression. 
Third, it is computationally efficient to implement the transform, due to the hierarchical and local nature of the computations. 

For some applications, the reconstruction process starts with noisy transform coefficients, due  
to communication noise, quantization error, or other types of processing. In this case, Do and Vetterli \cite{vetterli_pyr_short,vetterli_pyr_long} show that Burt and Adelson's single-level synthesis operator is not optimal in terms of the reconstruction error. The single-level analysis operator that transforms a coarse approximation $\mathbf{x}^{(j)}$ into the next coarse approximation $\mathbf{x}^{(j+1)}$ and prediction error $\mathbf{y}^{(j)}$ is a \emph{frame operator} (see, e.g., \cite{frames}). Frame operators have an infinite number of left inverses, but the one that minimizes the reconstruction error is the pseudoinverse of the analysis operator at each level of the pyramid. In Section \ref{Se:lp_synthesis}, we also use the pseudoinverse of the analysis operator as the synthesis operator at each level of our Laplacian pyramid for signals on graphs. Thus, we defer a more precise mathematical treatment of the optimal reconstruction until then.

\subsection{The Analysis Operator for a Single Level of the 
Pyramid for Signals on Graphs} \label{Se:LP_Analysis}
We now extend the classical Laplacian pyramid to a 
pyramid transform for signals on graphs. At each level of the pyramid, the inputs are the current coarse approximation, $\mathbf{x}^{(j)} \in \Rbb^{N^{(j)}}$, of the original signal, and the current graph Laplacian $\L^{(j)} \in \Rbb^{N^{(j)}} \times \Rbb^{N^{(j)}}$. We use the polarity of the largest eigenvector of $\L^{(j)}$, as described in Section \ref{Se:down}, to select the vertices of the graph on which to form the reduced graph at resolution scale $j+1$. We denote the output of the vertex selection process by $\mathbf{m}^{(j)} \in \Rbb^{N^{(j)}}$, where
\begin{align*}
{m}^{(j)}(i) =
\begin{cases}
\vspace{-.05in}
~1, & \hbox{ if vertex $i$ is selected to be included in the} \\
& \hbox{ reduced graph} \\
~0, & \hbox{ otherwise}
\end{cases}.
\end{align*}

The vertex selection vector $\mathbf{m}^{(j)}$ is used both to reduce the graph and to define the downsampling and 
interpolation operators. For graph reduction, we use Kron reduction followed by spectral sparsification, as 
discussed in Section \ref{Se:graph_reduction}. The graph Laplacian of the reduced graph is given by
\begin{align*}
&\L^{(j+1)}= \S\left(\K\left(\L^{(j)}, \mathbf{m}^{(j)}\right)\right) \\
&=\S\left(\L_{\V_1^{(j)},\V_1^{(j)}}^{(j)}-\L_{\V_1^{(j)},\V_1^{{(j)}^c}}^{(j)}\left[\L_{\V_1^{{(j)}^c},\V_1^{{(j)}^c}}^{(j)}\right]^{-1}\L_{\V_1^{{(j)}^c},\V_1^{(j)}}^{(j)}\right),
\end{align*}
where $\V_1^{(j)}:=\left\{i \in \{1,2,\ldots,N^{(j)}\}: {m}^{(j)}(i)=1\right\}$ and $\S$ is the spectral sparsification operator. The reduced graph $\L^{(j+1)}$ and the vertex selection vector $\mathbf{m}^{(j)}$, both of which need to be stored for reconstruction, are the first two outputs of the transform at each level of the pyramid.

The other two outputs are the coarse approximation vector $\mathbf{x}^{(j+1)}$ at the coarser resolution scale $j+1$, and the next prediction error vector $\mathbf{y}^{(j)}$. 
The course approximation vector is computed in
a similar fashion as in the classical Laplacian pyramid, with the filtering and downsampling
operators replaced by their graph analogs. The downsampling operator $\downarrow$ corresponds to matrix multiplication by the matrix $\mathbf{S}_d^{(j)} \in \Rbb^{N^{(j+1)}} \times \Rbb^{N^{(j)}}$, where
$\mathbf{S}_d^{(j)}:=\left[\mathbf{I}_{N^{(j)}}\right]_{\V_1^{(j)},\V^{(j)}}$
and 
$\mathbf{I}_{N^{(j)}}$ is an $N^{(j)} \times N^{(j)}$ identity matrix.
For example, if $\mathbf{m}^{(j)}=(1, 1, 0, 0, 1, 0)^{\transpose}$, then
\begin{align*}
\mathbf{S}_d^{(j)}=\left[
\begin{array}{cccccc}
1 & 0 & 0 & 0 & 0 & 0 \\
0 & 1 & 0 & 0 & 0 & 0 \\
0 & 0 & 0 & 0 & 1 & 0
\end{array}
\right].
\end{align*}
Rather than upsample and filter with another lowpass filter $G$ as in the classical pyramid, we compute the prediction error with the spline interpolation method of Section \ref{Se:interpolation}. While this adds some additional computational cost, our experiments showed that this change significantly improves the predictions, and in turn the sparsity of the transform coefficients. 
So the single-level analysis operator $\mathbf{T}_a^{(j)}: \Rbb^{N^{(j)}}\rightarrow \Rbb^{N^{(j)}+N^{(j+1)}}$ is  
given by
\begin{align}\label{Eq:graph_analysis_op}
\mathbf{T}_a^{(j)} \mathbf{x}^{(j)}&:=\left[
\begin{array}{c}
\mathbf{S}_d^{(j)} \mathbf{H}^{(j)} \\
\mathbf{I}_{N^{(j)}}-\mathbf{\Phi}_{\V_1}^{(j)}\left(\mathbf{\Phi}_{\V_1,\V_1}^{(j)}\right)^{-1}\mathbf{S}_d^{(j)}\mathbf{H}^{(j)}
\end{array}
\right]\mathbf{x}^{(j)} \nonumber \\&~=\left[
\begin{array}{c}
\mathbf{x}^{(j+1)} \\
\mathbf{y}^{(j)}
\end{array}
\right].
\end{align}
In \eqref{Eq:graph_analysis_op},  the filtering operator is
$\mathbf{H}^{(j)}=\mathbf{U}^{(j)}[\hat{h}(\mathbf{\Lambda}^{(j)})] \mathbf{U}^{{(j)}^{*}}$.

A schematic representation of a single level of the  
pyramid for signals on graphs is shown in Figure \ref{Fig:lp}(b).
Note that while we use the polarity of the largest eigenvector of the graph Laplacian as a vertex selection method, the sparsified Kron reduction as a graph reduction method, the eigenvectors of the graph Laplacian as a filtering basis, and the spline interpolation method with translated Green's functions, other vertex selection, graph reduction, filtering, and interpolation methods can be substituted into these blocks without affecting the 
perfect reconstruction property of the scheme.

\subsection{The Multilevel Pyramid}
The multilevel 
pyramid for signals on graphs is shown in Figure \ref{Fig:lp}(c). As in the classical case, we iterate the single-level analysis on the downsampled output of the lowpass channel. 
In \eqref{Eq:graph_analysis_op}, the filtering operators $\mathbf{H}^{(j)}$ 
necessarily depend on the resolution scale $j$ through their dependence on the eigenvectors and eigenvalues of $\L^{(j)}$. However, the filter 
$\hat{h}: [0, \lambda_{\max}^{(0)}] \rightarrow \Rbb$ may be fixed across all levels of the pyramid due to property \eqref{Eq:Kron_spectrum} of the Kron reduction, which guarantees that the spectrum of ${\L}^{(j)}$ is contained in $[0, \lambda_{\max}^{(0)}]$ for all $j$.\footnote{If we include spectral sparsification in the graph reduction step, then the spectrum of $\L^{(j)}$ is not guaranteed to be contained in $[0,\lambda_{\max}^{(0)}]$; however, for large graphs, the results of \cite{spielman_srivastava} limit the extent to which the maximum graph Laplacian eigenvalue can increase beyond $\lambda_{\max}^{(0)}$. In any case, we usually define the filters on the entire positive real line, in which case there is no issue with the length of the spectrum increasing at successive levels of the pyramid.}

\begin{figure*}[htb]
\centering
\centerline{\includegraphics[width=\linewidth]{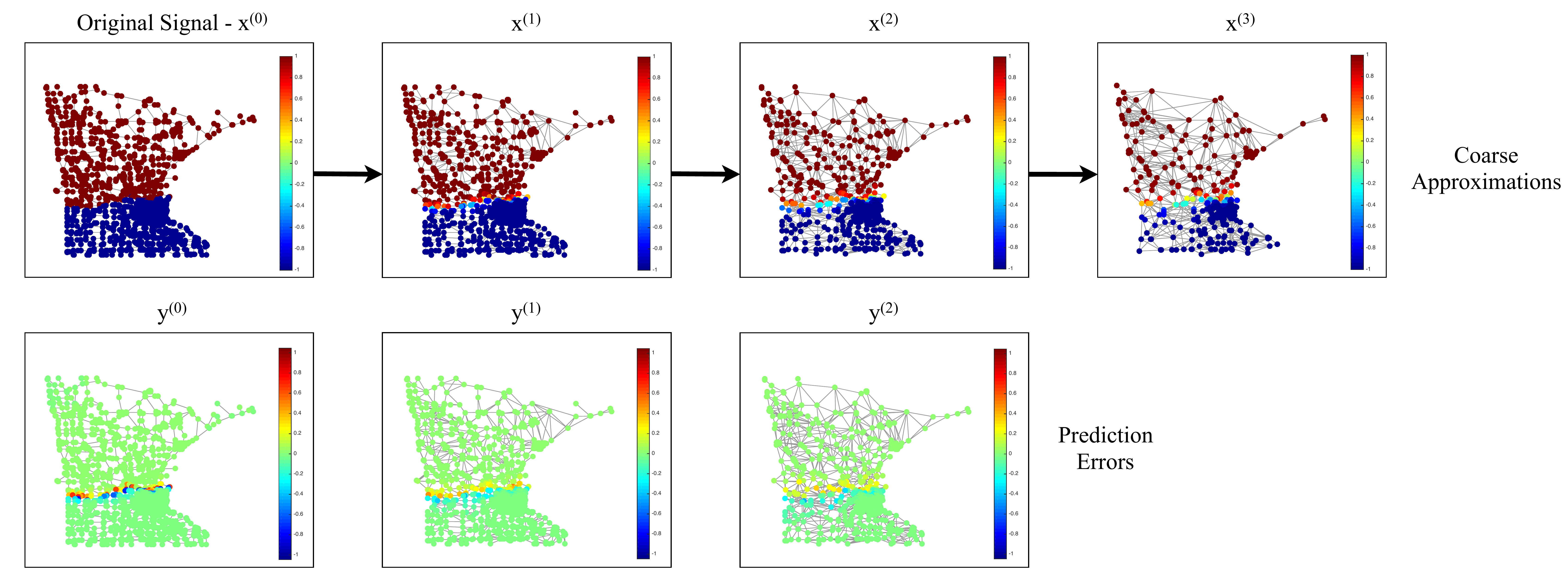}}
 \caption {Three-level pyramid analysis of a piecewise-constant signal on the Minnesota road network of \cite{gleich}. The number of vertices is reduced in successive coarse approximations from $N=2642$ to $1329$ to $676$ to $334$.  The overall redundancy factor of the transform in this case is 1.89.}
  \label{Fig:lp_example}
\end{figure*}
\begin{figure*}[htb]
\centering
\centerline{\includegraphics[width=\linewidth]{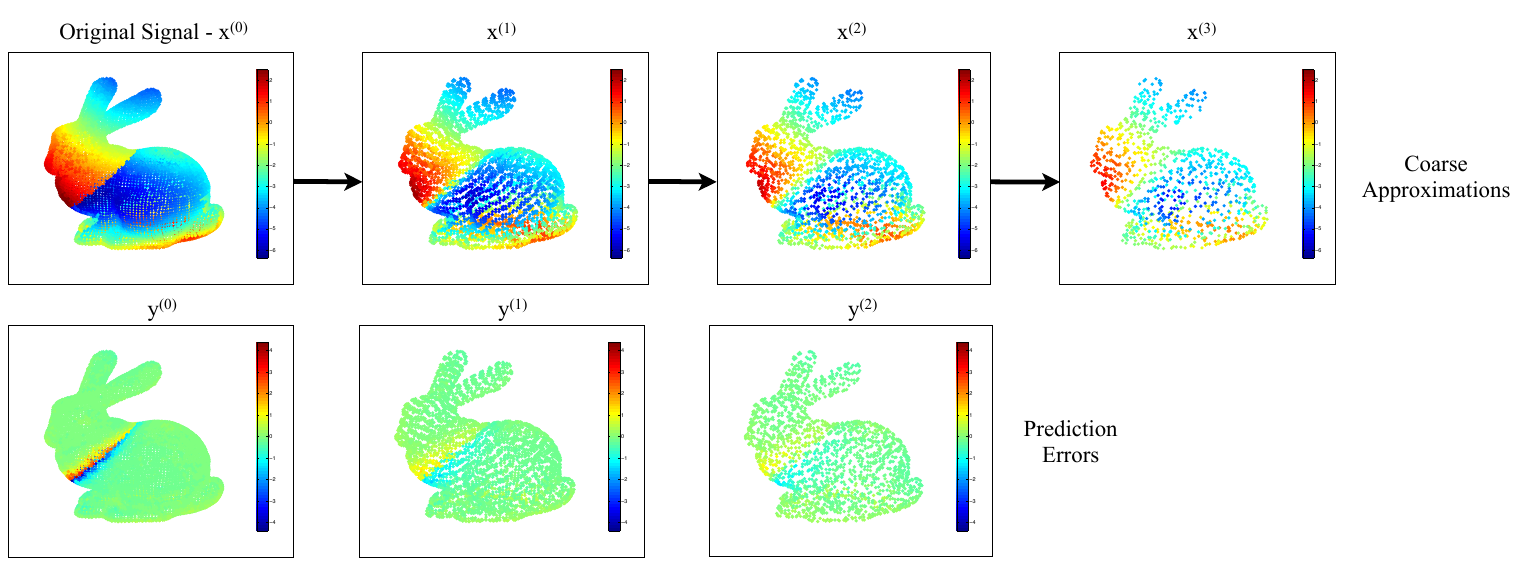}}
 \caption {Three-level pyramid analysis of a piecewise-smooth signal on the Stanford Bunny graph of \cite{bunny}. The number of vertices is reduced in successive coarse approximations from $N=8170$ to $4108$ to $2050$ to $1018$.  The overall redundancy factor of the transform in this case is 1.88.}
  \label{Fig:lp_bunny}
\end{figure*}

\subsection{Synthesis Operators for a Single Level and Optimal Reconstruction} \label{Se:lp_synthesis}
Just as is the case in the classical Laplacian pyramid, for any choices of filter 
$\hat{h}(\cdot)$, 
we can reconstruct $\mathbf{x}^{(j)}$ perfectly from $\mathbf{x}^{(j+1)}$ and $\mathbf{y}^{(j)}$ with a single-level synthesis operator, $\mathbf{T}_s^{(j)}:\Rbb^{N^{(j)}+N^{(j+1)}} \rightarrow \Rbb^{N^{(j)}}$, given by
\begin{align}\label{Eq:synthesis_op}
\mathbf{T}_s^{(j)}\left[
\begin{array}{c}
\mathbf{x}^{(j+1)} \\
\mathbf{y}^{(j)}
\end{array}
\right]:=\left[\mathbf{\Phi}_{\V_1}^{(j)}\left(\mathbf{\Phi}_{\V_1,\V_1}^{(j)}\right)^{-1} ~~\mathbf{I}_{N^{(j)}} \right]
\left[
\begin{array}{c}
\mathbf{x}^{(j+1)} \\
\mathbf{y}^{(j)}
\end{array}
\right].
\end{align}
Note that in order to apply the synthesis operator $\mathbf{T}_s^{(j)}$, we need access to both $\L^{(j)}$ and $\mathbf{m}^{(j)}$, which are used for the interpolation.

It is straightforward to check that
$ \mathbf{T}_s^{(j)} \mathbf{T}_a^{(j)}=\mathbf{I}_{N^{(j)}}$,
guaranteeing perfect reconstruction. However, once again analogously to the classical Laplacian pyramid, when noise is introduced into the stored coefficients $\mathbf{x}^{(j+1)}$ and $\mathbf{y}^{(j)}$, we would like to reconstruct with the pseudoinverse of $\mathbf{T}_a^{(j)}$:
\begin{align}\label{Eq:pseudo}
\mathbf{T}_a^{{(j)}^{\dagger}}:=\left(\mathbf{T}_a^{{(j)}^{*}}\mathbf{T}_a^{(j)}\right)^{-1}\mathbf{T}_a^{{(j)}^{*}}.
\end{align}
If $\tilde{\mathbf{x}}^{(j+1)}$ and $\tilde{\mathbf{y}}^{(j)}$ are the noisy versions of $\mathbf{x}^{(j+1)}$ and $\mathbf{y}^{(j)}$, respectively, then ${\breve{\mathbf{x}}}^{(j)}=\mathbf{T}_a^{{(j)}^{\dagger}}\left[
\tilde{\mathbf{x}}^{(j+1)} \atop
\tilde{\mathbf{y}}^{(j)}
\right]$ minimizes the reconstruction error. That is,
\begin{align}\label{Eq:pseudo_applied}
\mathbf{T}_a^{{(j)}^{\dagger}}\left[
\tilde{\mathbf{x}}^{(j+1)} \atop
\tilde{\mathbf{y}}^{(j)}
\right] = \argmin_{\mathbf{x} \in \Rbb^{N^{(j)}}}
{\left|\left|\mathbf{T}_a^{(j)} \mathbf{x} -\left[
\tilde{\mathbf{x}}^{(j+1)} \atop
\tilde{\mathbf{y}}^{(j)}
\right] \right|\right|}_2.
\end{align}
For large graphs, rather than form and compute the matrix inverse in \eqref{Eq:pseudo}, 
it may be more computationally efficient to approximate the left-hand side of \eqref{Eq:pseudo_applied} with Landweber iteration \cite[Theorem 6.1]{engl}.

To summarize, for a multilevel 
pyramid for signals on graphs that has $J$ resolution levels, the reconstruction process begins with knowledge of $\mathbf{m}^{(0)}$, $\mathbf{m}^{(1)}$, $\ldots$, $\mathbf{m}^{(J-1)}$, $\L^{(0)}$, $\L^{(1)}$, $\ldots$, $\L^{(J-1)}$, $\mathbf{y}^{(0)}$, $\mathbf{y}^{(1)}$, $\ldots$, $\mathbf{y}^{(J-1)}$, and $\mathbf{x}^{(J)}$. We first use $\mathbf{m}^{(J-1)}$ and $\L^{(J-1)}$ to apply $\mathbf{T}_a^{{(J-1)}^{\dagger}}$ to $\left[{\mathbf{x}}^{(J)} \atop
{\mathbf{y}}^{(J-1)} \right]$. The result is $\breve{\mathbf{x}}^{(J-1)}$. We then use $\mathbf{m}^{(J-2)}$ and $\L^{(J-2)}$ to apply $\mathbf{T}_a^{{(J-2)}^{\dagger}}$ to $\left[{\breve{\mathbf{x}}}^{(J-1)} \atop
{\mathbf{y}}^{(J-2)} \right]$ in order to compute $\breve{\mathbf{x}}^{(J-2)}$. We iterate this process $J$ times until we finally compute $\breve{\mathbf{x}}^{(0)}$, which, if no noise has been introduced into the transform coefficients, is equal to the original signal ${\mathbf{x}}^{(0)}$.

\subsection{Illustrative Examples}
In Figure \ref{Fig:lp_example}, we use the 
proposed pyramid transform to analyze a piecewise-constant signal (the sign of the Fiedler vector $\mathbf{u}_1$) on the Minnesota road network of \cite{gleich}. 
After each Kron reduction at stage $j$, we use the spectral sparsification of Algorithm \ref{Al:sparsification} with $Q$ set to the integer closest to $4N^{(j)}\log(N^{(j)})$.
The graph spectral filter
is 
$\hat{h}(\lambda_{\l}):=\frac{0.5}{0.5+\lambda_{\l}}$, and we take $\epsilon=0.005$ to form the regularized Laplacian for interpolation. 
We see that the prediction error coefficients are extremely sparse, and those with 
non-zero
magnitudes are concentrated around the discontinuity in the graph signal.

In Figure \ref{Fig:lp_bunny}, we apply the transform to a piecewise smooth signal on the Stanford Bunny graph \cite{bunny} with 8170 vertices. 
The signal is comprised of two different polynomials of the coordinates: $\bar{x}^2+\bar{y}-2\bar{z}$ on the head and ears of the bunny, and $\bar{x}-\bar{y}+3\bar{z}^2-5$ on the lower body, where $(\bar{x},\bar{y},\bar{z})$ are the physical coordinates of each vertex.
Since the graph is larger and the graph Laplacian spectrum is wider than the previous example, we take $Q$ to be the integer closest to $16N^{(j)}\log(N^{(j)})$, the lowpass filter to be $\hat{h}(\lambda_{\l}):=\frac{5}{5+\lambda_{\l}}$, and $\epsilon=0.05$ for interpolation.

\begin{figure}[tb]
\centering
\begin{minipage}[b]{.32 \linewidth}
   \centering
   \centerline{\includegraphics[width=\linewidth]{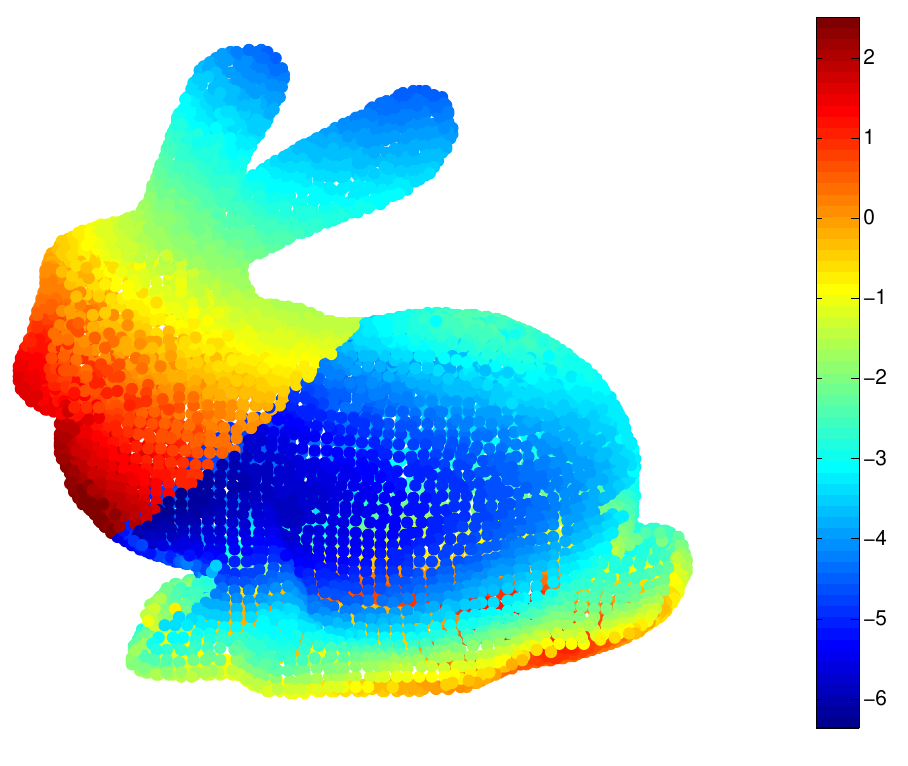}}
\centerline{\small{(a)}}
\end{minipage} 
\begin{minipage}[b]{.32 \linewidth}
   \centering
   \centerline{\includegraphics[width=\linewidth]{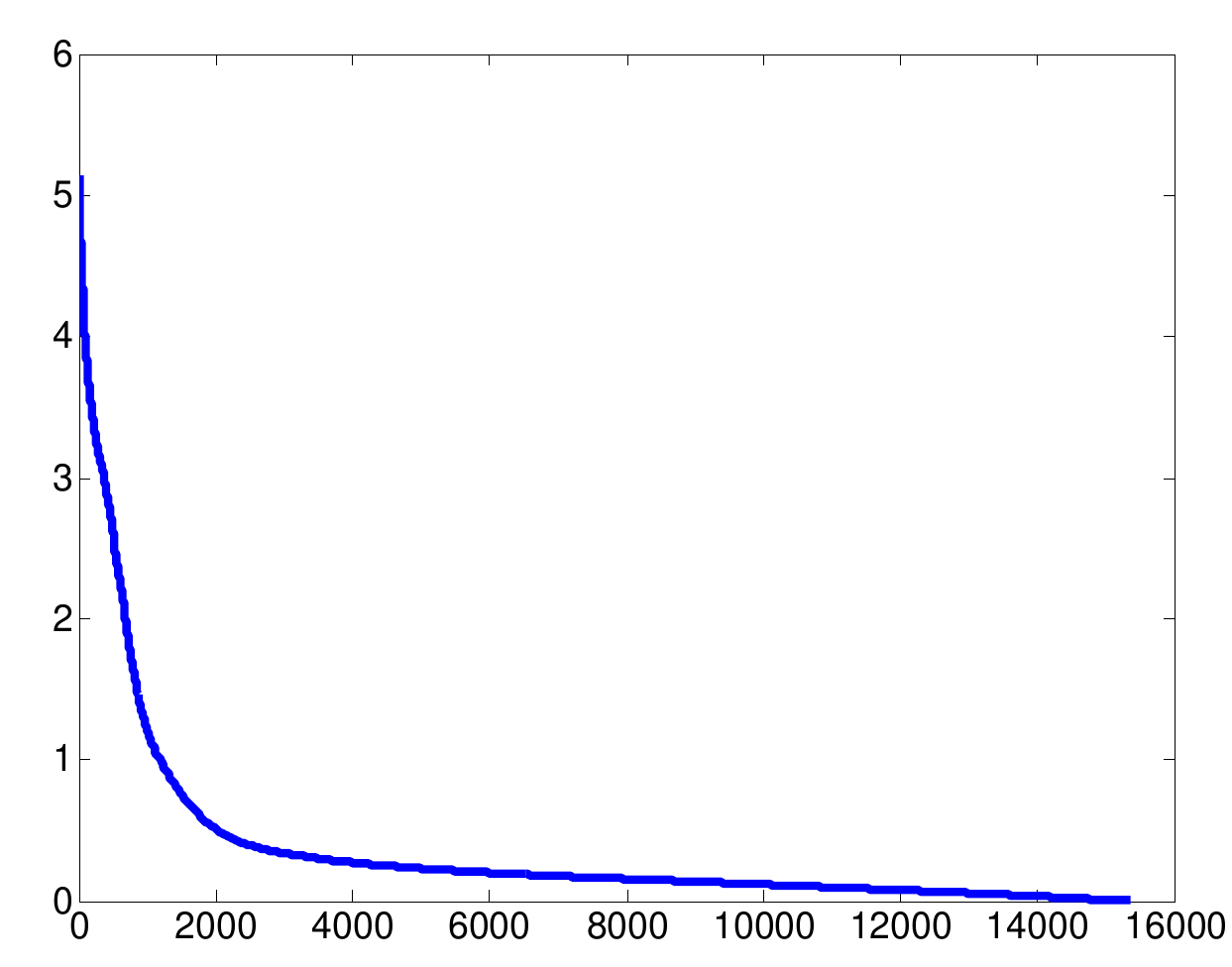}}
\centerline{\small{(b)}}
\end{minipage}
\begin{minipage}[b]{.32 \linewidth}
   \centering
   \centerline{\includegraphics[width=\linewidth]{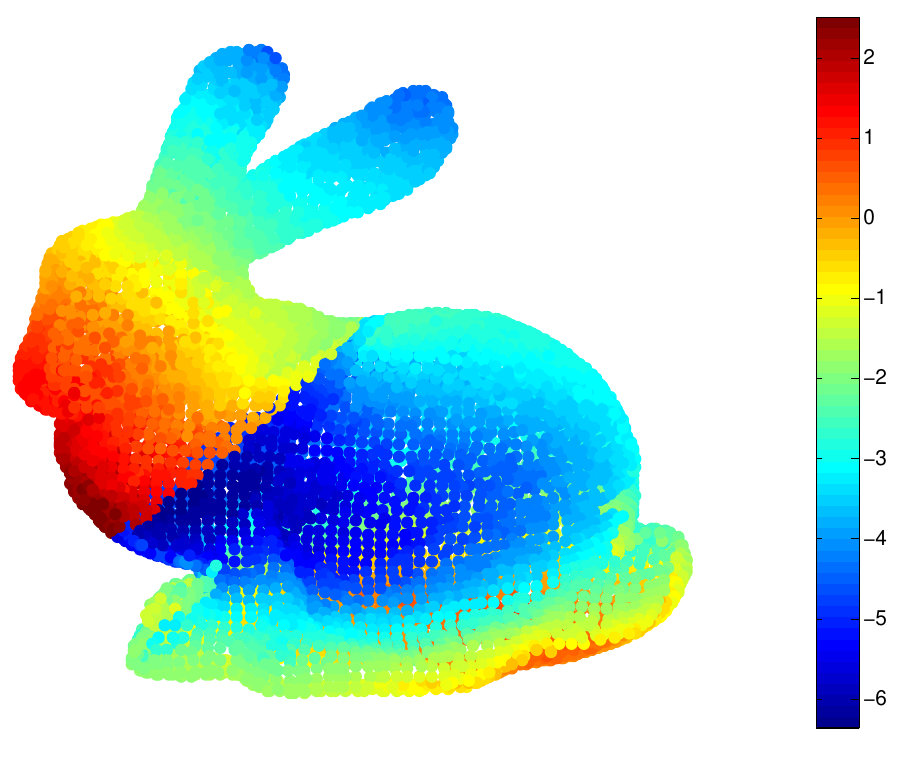}}
\centerline{\small{(c)}}
\end{minipage}
\caption {Compression example. (a) The original piecewise-smooth signal with a discontinuity on the Stanford bunny \cite{bunny}. (b) The sorted magnitudes of the 15346 pyramid transform coefficients. (c) The reconstruction from the 2724 coefficients with the largest magnitudes, using the least squares synthesis \eqref{Eq:pseudo}.}
  \label{Fig:bunny_compressions}
\end{figure}

As a compression example, we hard threshold the 15346 pyramid transform coefficients of the signal, keeping only the 2724 with the largest magnitudes (approximately 1/3 the size of the original signal) and setting the rest to 0. Reconstruction of the signal from the thresholded coefficients yields an error $\frac{||\mathbf{f}_{\textrm{reconstruction}}-\mathbf{f}||_2}{||\mathbf{f}||_2}$ of 0.145 with the direct synthesis \eqref{Eq:synthesis_op}, and an error of 0.086 with the least squares synthesis \eqref{Eq:pseudo}. Figure \ref{Fig:bunny_compressions} shows the original signal, the sorted magnitudes of the  transform coefficients, and the reconstruction with the least squares synthesis operator.

\subsection{Comparison with Other Transforms}
A thorough comparison of our proposed pyramid transform to other multiscale transforms for signals on graphs investigating which transforms work best on which types of graph signals in which signal processing tasks is beyond the scope of this work. However, in this section, we present a few illustrative numerical denoising and compression experiments with our proposed pyramid transform and other multiscale transforms for graph signals.

First, one qualitative difference between the proposed transform and many other multiscale transforms for graph signals is that our transform outputs both a multiresolution of graphs and a graph signal multiresolution residing on that graph multiresolution. While less important for numerical graph signal processing and machine learning tasks, this feature is especially beneficial for multiscale visualization of graph signals (e.g., the sequence of coarse approximations in the top row of Figure \ref{Fig:lp_bunny}). A few other transforms such as the critically-sampled spline wavelet filter banks \cite{ekambaram_globalsip} share this feature, but many others such as the spectral graph wavelets \cite{sgwt} and the spatial graph wavelets \cite{Crovella2003} do not yield such a sequence of coarse approximations at different resolution levels. 
 
Second, the dictionary atoms we construct tend to be jointly localized in the vertex domain and the graph spectral domain. The hope is that such atoms are able to sparsely represent various classes of graph signals, such as those that are piecewise smooth with discontinuities. Unfortunately, there is little theory to date relating mathematical classes of graph signals to the sparsity of particular transform coefficients. One initial exploration into this line of investigation for the case of spectral graph wavelets is presented in \cite{ricaud_SPIE_2013}. We conduct a few numerical experiments below to explore the decay of the transform coefficients for different types of signals.

\subsubsection{Denoising}

\begin{figure}[tb]
\centering
\begin{minipage}[b]{.32 \linewidth}
   \centering
   \centerline{\includegraphics[width=\linewidth]{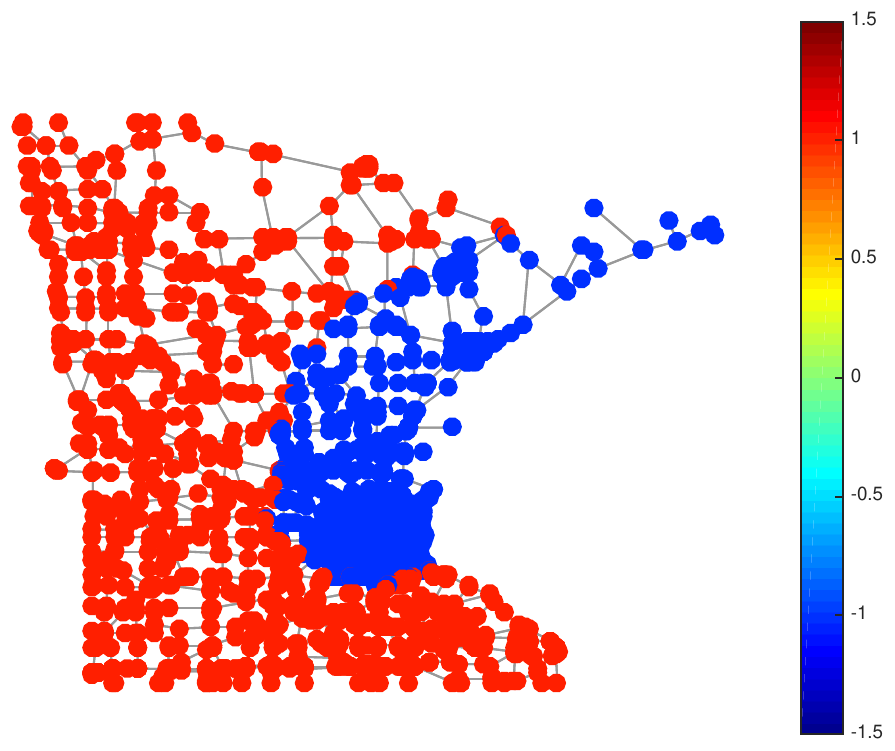}}
\centerline{\small{(a)}}
\end{minipage} 
\begin{minipage}[b]{.32 \linewidth}
   \centering
   \centerline{\includegraphics[width=\linewidth]{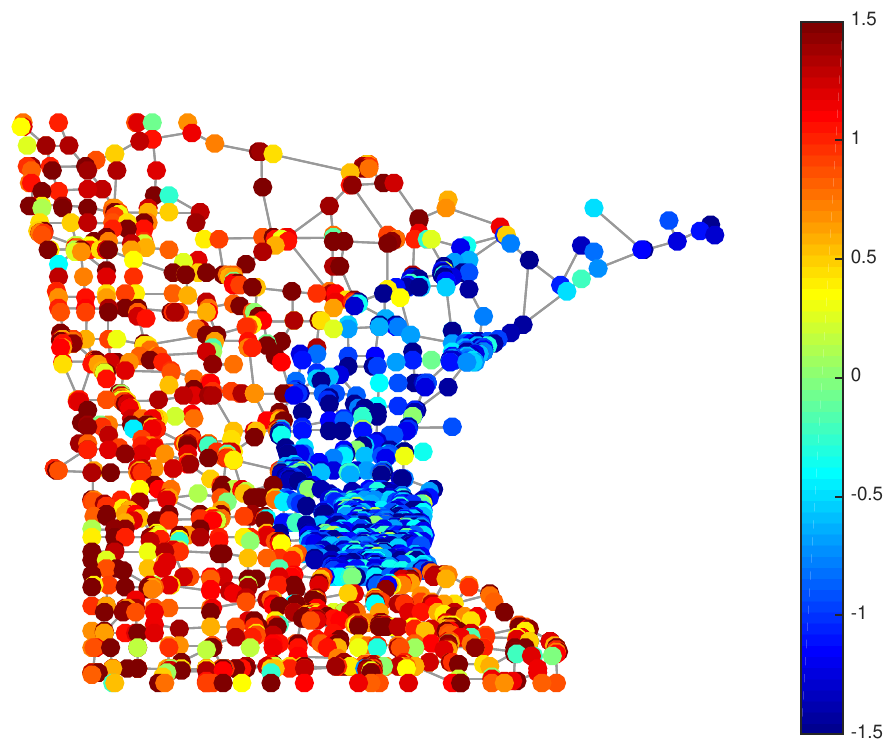}}
\centerline{\small{(b)}}
\end{minipage}
\begin{minipage}[b]{.32 \linewidth}
   \centering
   \centerline{\includegraphics[width=\linewidth]{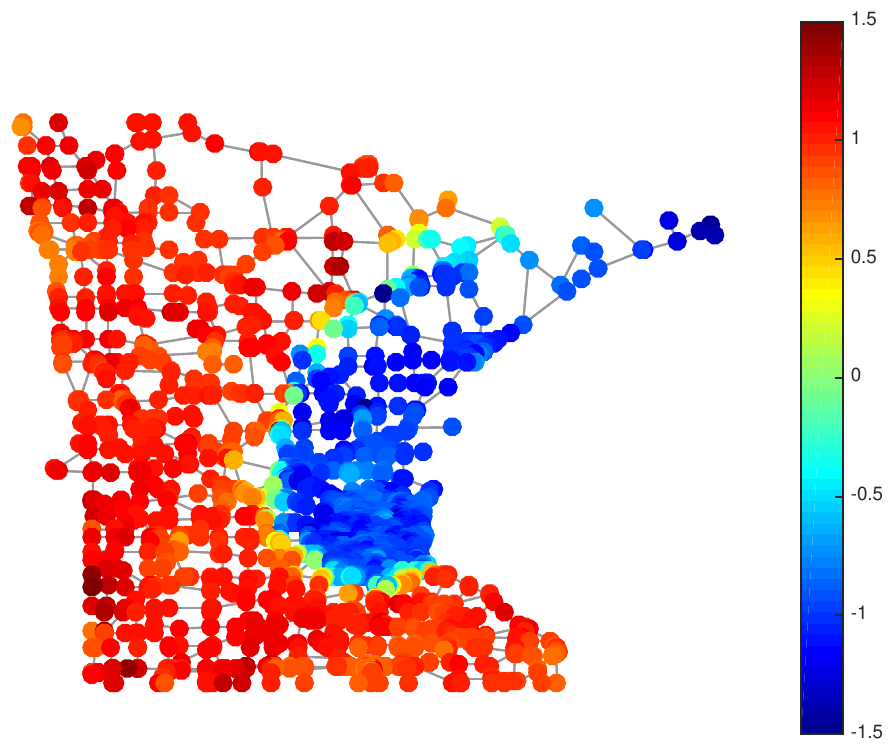}}
\centerline{\small{(c)}}
\end{minipage}
\caption {Denoising example. (a) Piecewise constant signal on the Minnesota graph. (b) Noisy observation with $\sigma=\frac{1}{2}$. (c) Denoised signal reconstructed after hard thresholding the prediction errors of a two-level pyramid transform.}
  \label{Fig:denoising}
\end{figure}

We start by repeating the denoising experiment of \cite[Section VI.D]{sakiyama} on the Minnesota road network of \cite{gleich}.\footnote{To replicate the experiment with our transform, we leveraged the MATLAB code of \cite{sakiyama}, which is  publicly available at http://tanaka.msp-lab.org/software.} We add white Gaussian noise with varying standard deviations to the piecewise-constant reference signal from \cite{narang_bipartite_prod} shown in Figure \ref{Fig:denoising}(a). As in \cite{sakiyama}, we compute the pyramid transform coefficients, hard threshold the prediction errors at a threshold of $3\sigma$, and reconstruct a denoised signal (shown in Figure \ref{Fig:denoising}(c) for the case of $\sigma=\frac{1}{2}$) from the coarse approximation coefficients and the thresholded prediction error coefficients. We use the same parameters as in Figure \ref{Fig:lp_example} to construct the graph multiresolution. We take the graph spectral filters to be
$\hat{h}(\lambda_{\l}):=\frac{1}{1+\lambda_{\l}}$, and take $\epsilon=0.005$ to form the regularized Laplacian for interpolation. In Table \ref{Ta:denoising}, we compare the denoising results for a one-level and two-level transform to a reference range of results attained by different transforms in \cite{sakiyama}, 
including variants of the biorthogonal graph filter banks of \cite{narang_bior_filters}, the oversampled graph filter banks of \cite{sakiyama}, and the spectral graph wavelet transform of \cite{sgwt}. Qualitatively comparing the denoised signal in Figure \ref{Fig:denoising}(c) to its counterparts in \cite[Fig. 15]{sakiyama}, it appears that our denoised signal is smoother with respect to the graph structure, with more of the errors clustered around the discontinuity in the piecewise-constant signal.

\begin{table}[t]
\caption{Denoising Results 
on the Minnesota Traffic Graph: SNR (dB)} 
{\footnotesize
\tabcolsep=0.11cm
\begin{tabular}{c||c|c|c|c}
\cline{1-5}
$\sigma$ & ~~noisy~~ & 1-level pyramid & 2-level pyramid & reference range \\
\cline{1-5}
1/32 & 30.10 & 34.80 & 34.13 & 31.44~-~35.08  \\ 
\cline{1-5}
1/16 & 24.08 & 28.57 & 27.11 & 25.61~-~29.34  \\ 
\cline{1-5}
1/8 & 18.06 & 22.88 & 20.44 & 19.97~-~23.17  \\ 
\cline{1-5}
1/4 & 12.04 & 17.00 & 15.94 & 14.19~-~17.63  \\ 
\cline{1-5}
1/2 & 6.02 & 12.76 & 13.66 & ~8.50~-~12.31  \\ 
\cline{1-5}
1 & 0.00 & 7.94 & 10.64 & ~2.63~-~~8.82  \\ 
\hline \hline
Redundancy & - & 1.50 & 1.76 & ~1.00~-~~4.00 \\
\hline
\end{tabular}
}
\label{Ta:denoising}
\vspace{-.5cm}
\end{table}

\begin{figure*}[thb]
\centering
\hfill
\begin{minipage}[b]{.2 \linewidth}
   \centering
   \centerline{\includegraphics[width=\linewidth]{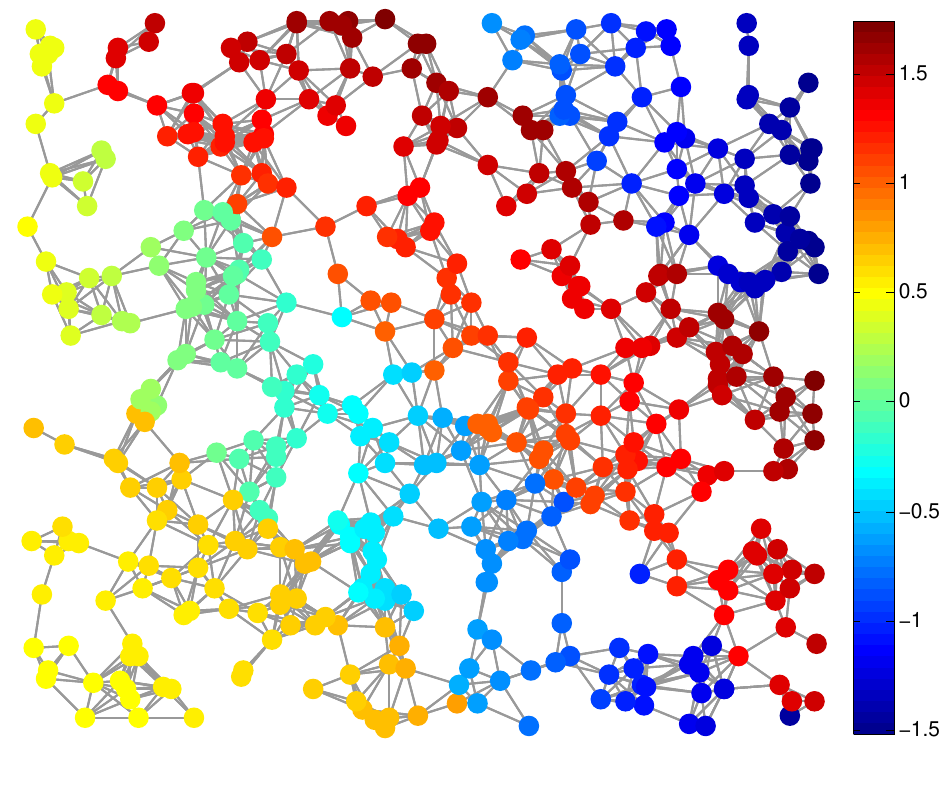}}
\centerline{\small{(a)}}
\end{minipage} 
\hfill
\begin{minipage}[b]{.21 \linewidth}
   \centering
   \centerline{\includegraphics[width=\linewidth]{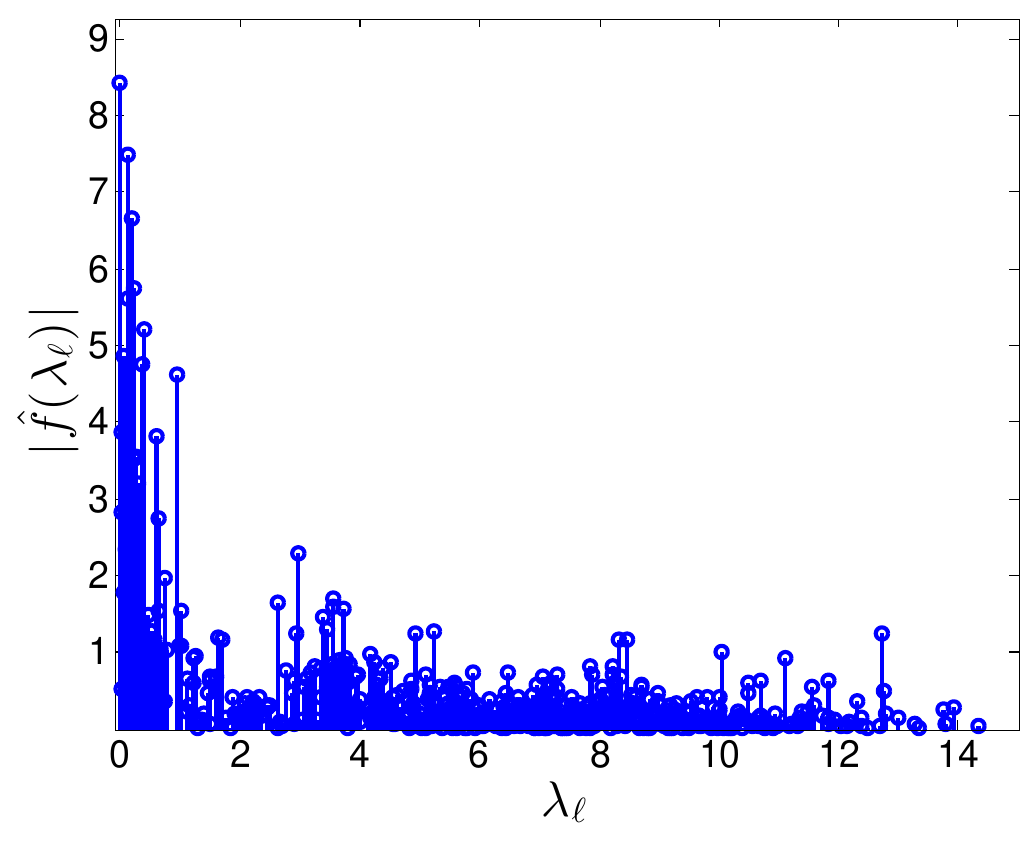}}
\centerline{\small{(b)}}
\end{minipage}
\hfill
\begin{minipage}[b]{.24 \linewidth}
   \centering
      \centerline{\includegraphics[width=\linewidth]{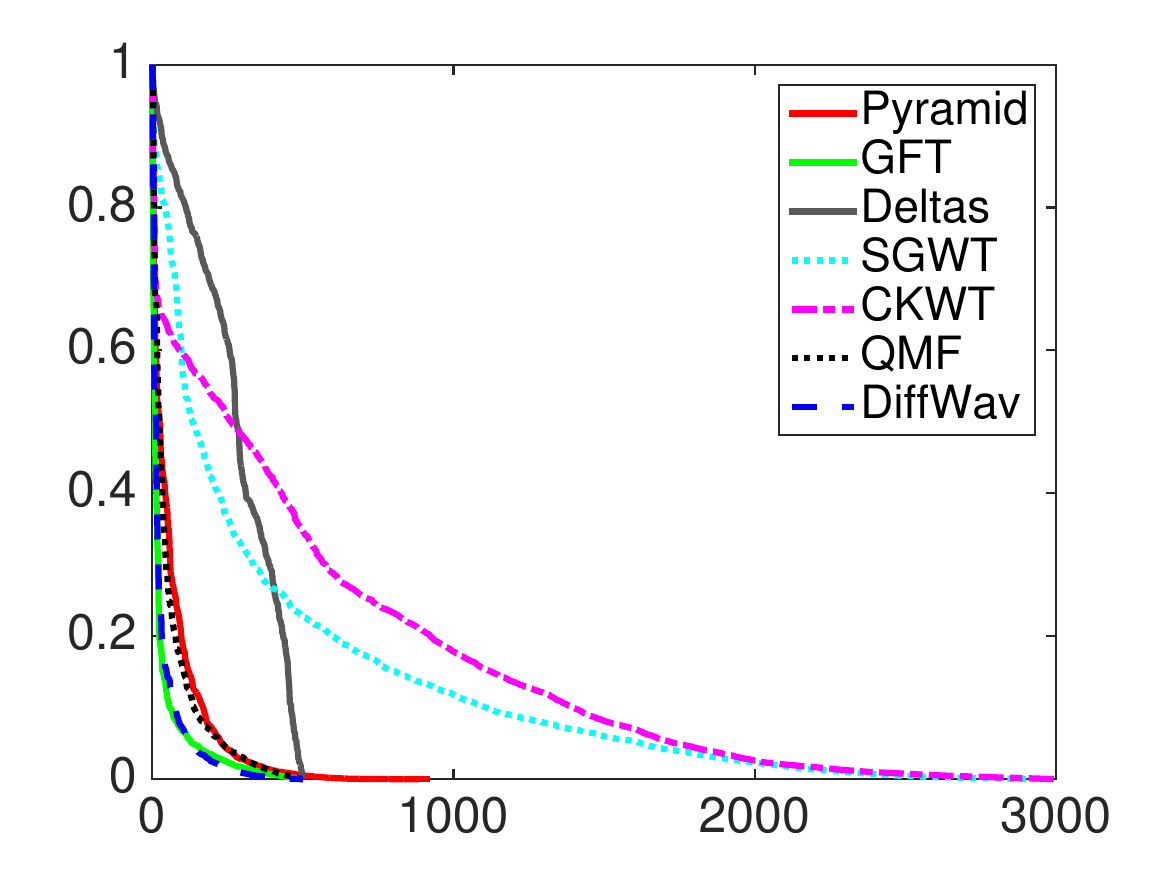}}
\centerline{\small{(c)}}
\end{minipage} 
\hfill
\begin{minipage}[b]{.24 \linewidth}
   \centering
   \centerline{\includegraphics[width=\linewidth]{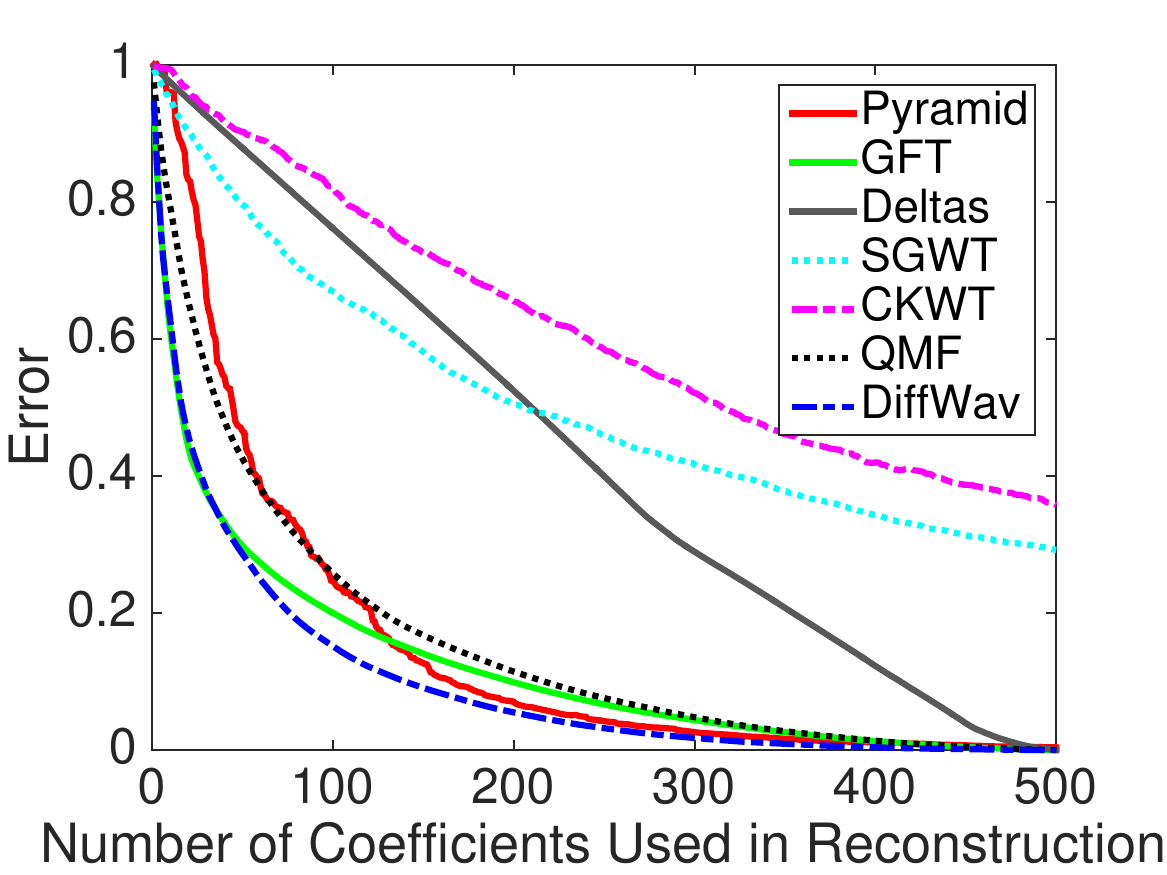}}
\centerline{\small{(d)}}
\end{minipage} 
\hfill
\caption {Compression example I. (a) A piecewise-smooth graph signal on a random sensor network with 500 vertices. (b) The same signal in the graph spectral domain. (c) The normalized sorted magnitudes of the transform coefficients. There are 500 coefficients each for the deltas, graph Fourier transform, QMF filter bank, and diffusion wavelets, 922 coefficients for the pyramid transform, and 3000 coefficients each for the spectral graph wavelets and spatial graph wavelets. (d) The reconstruction errors $\frac{||\mathbf{f}_{\textrm{reconstruction}}-\mathbf{f}||_2}{||\mathbf{f}||_2}$ resulting from hard thresholding each transform's coefficients, as a function of the number of coefficients used in the reconstruction.} 
  \label{Fig:compression}
\end{figure*}

\begin{figure*}[thb]
\centering
\hfill
\begin{minipage}[b]{.2 \linewidth}
   \centering
   \centerline{\includegraphics[width=\linewidth]{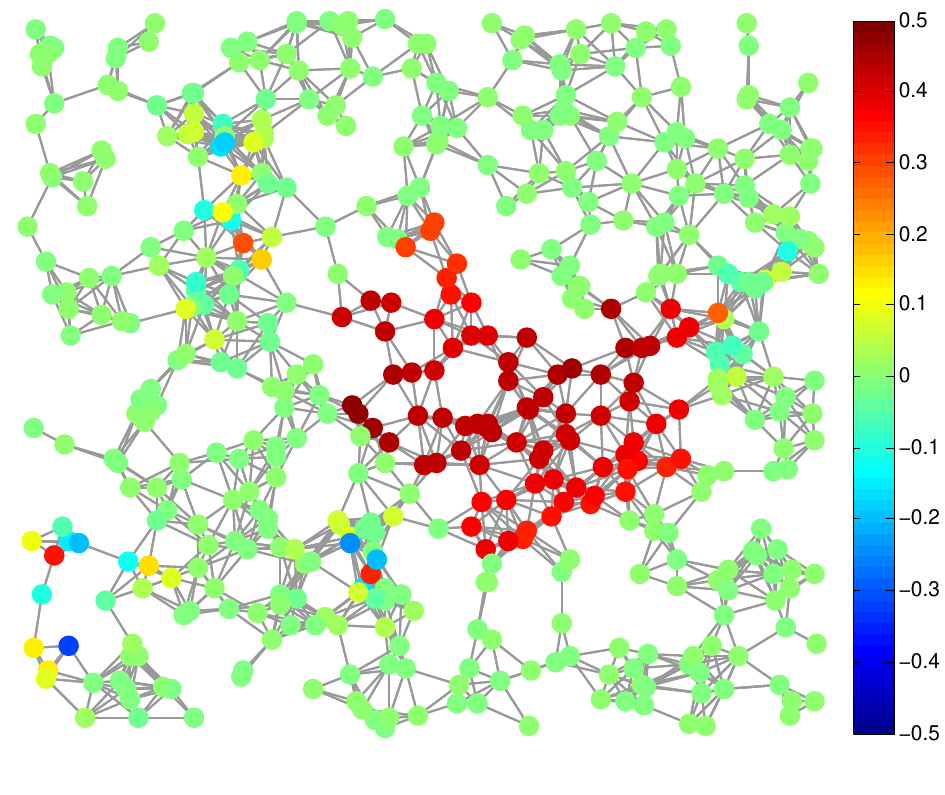}}
\centerline{\small{(a)}}
\end{minipage} 
\hfill
\begin{minipage}[b]{.215 \linewidth}
   \centering
   \centerline{\includegraphics[width=\linewidth]{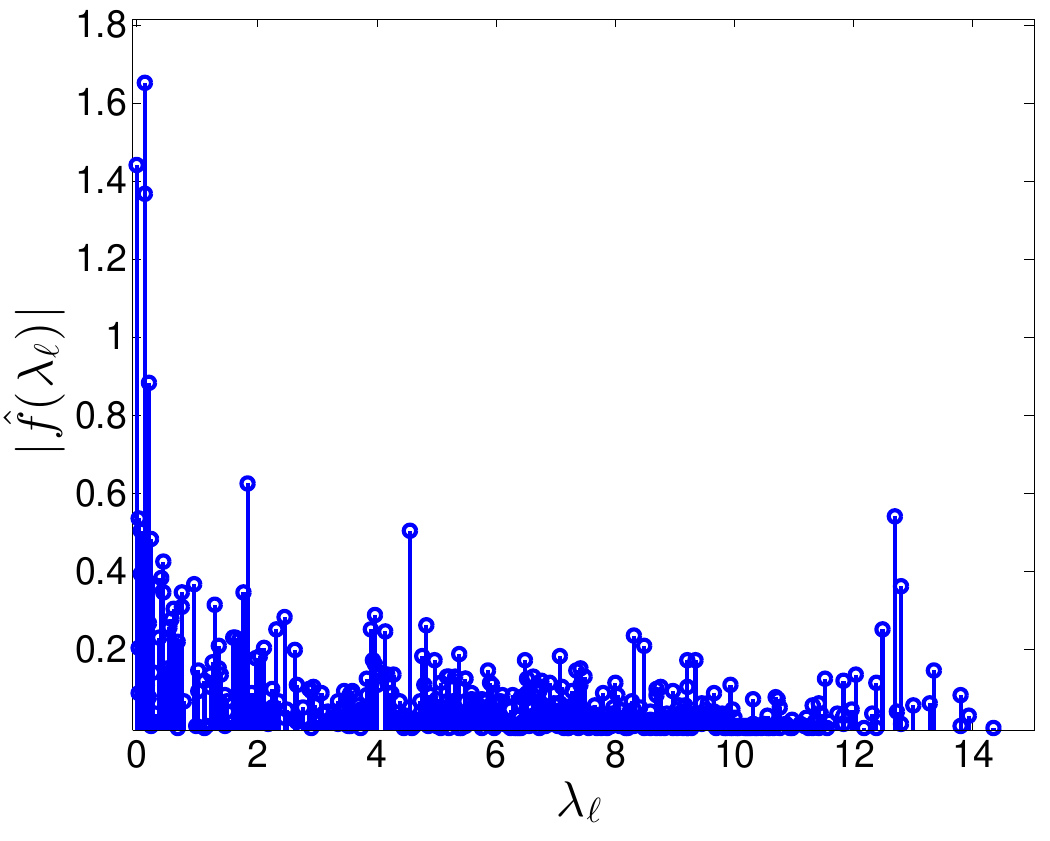}}
\centerline{\small{(b)}}
\end{minipage}
\hfill
\begin{minipage}[b]{.24 \linewidth}
   \centering
         \centerline{\includegraphics[width=\linewidth]{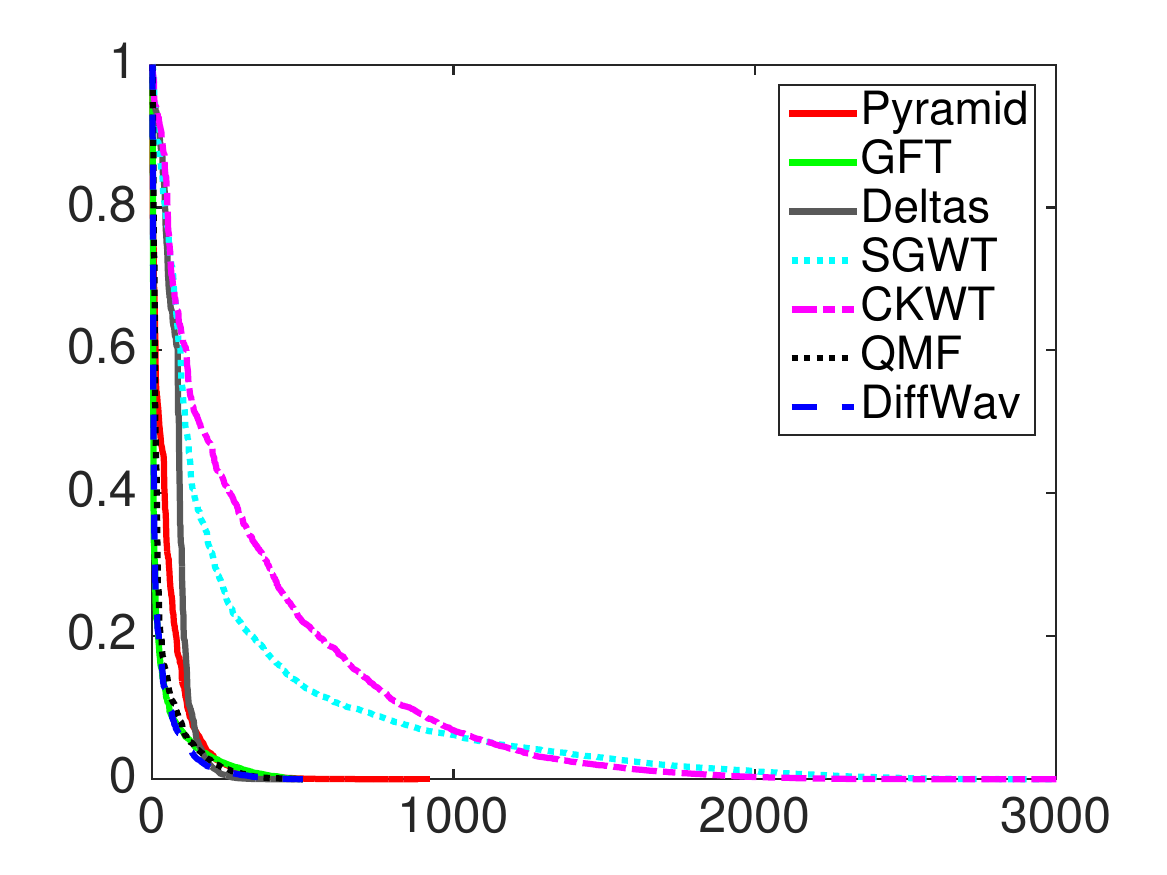}}
\centerline{\small{(c)}}
\end{minipage} 
\hfill
\begin{minipage}[b]{.24 \linewidth}
   \centering
      \centerline{\includegraphics[width=\linewidth]{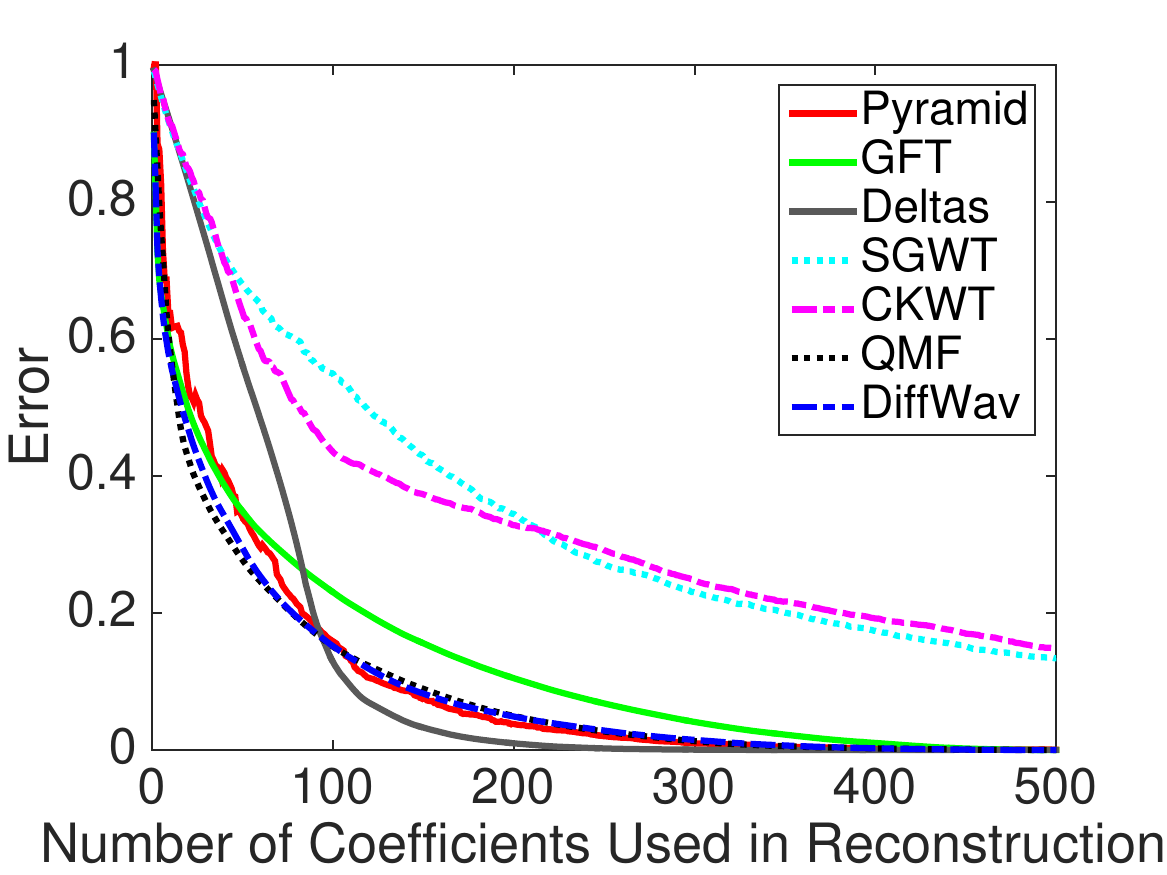}}
\centerline{\small{(d)}}
\end{minipage} 
\hfill
\caption {Compression example II. (a) A signal comprised of four different bandpass signals restricted to four different parts of the graph. (b) The same signal in the graph spectral domain. (c) The normalized sorted magnitudes of the transform coefficients. (d) The reconstruction errors $\frac{||\mathbf{f}_{\textrm{reconstruction}}-\mathbf{f}||_2}{||\mathbf{f}||_2}$.
}
  \label{Fig:compression2}
\end{figure*}

\begin{figure*}[thb]
\centering
\hfill
\begin{minipage}[b]{.2 \linewidth}
   \centering
   \centerline{\includegraphics[width=\linewidth]{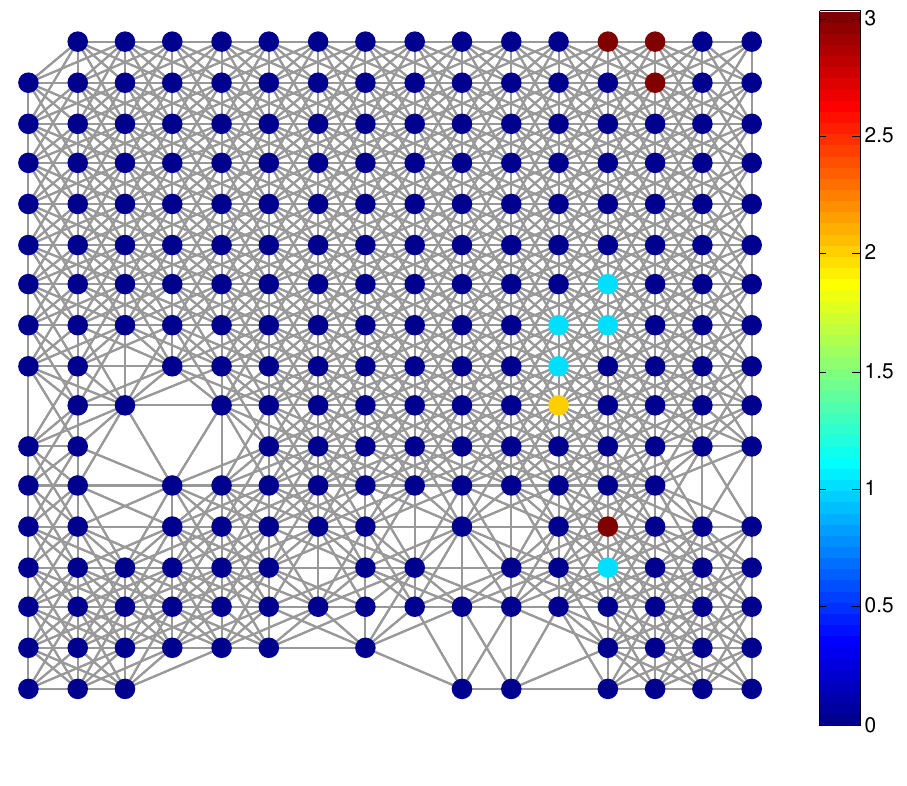}}
\centerline{\small{(a)}}
\end{minipage} 
\hfill
\begin{minipage}[b]{.215 \linewidth}
   \centering
   \centerline{\includegraphics[width=\linewidth]{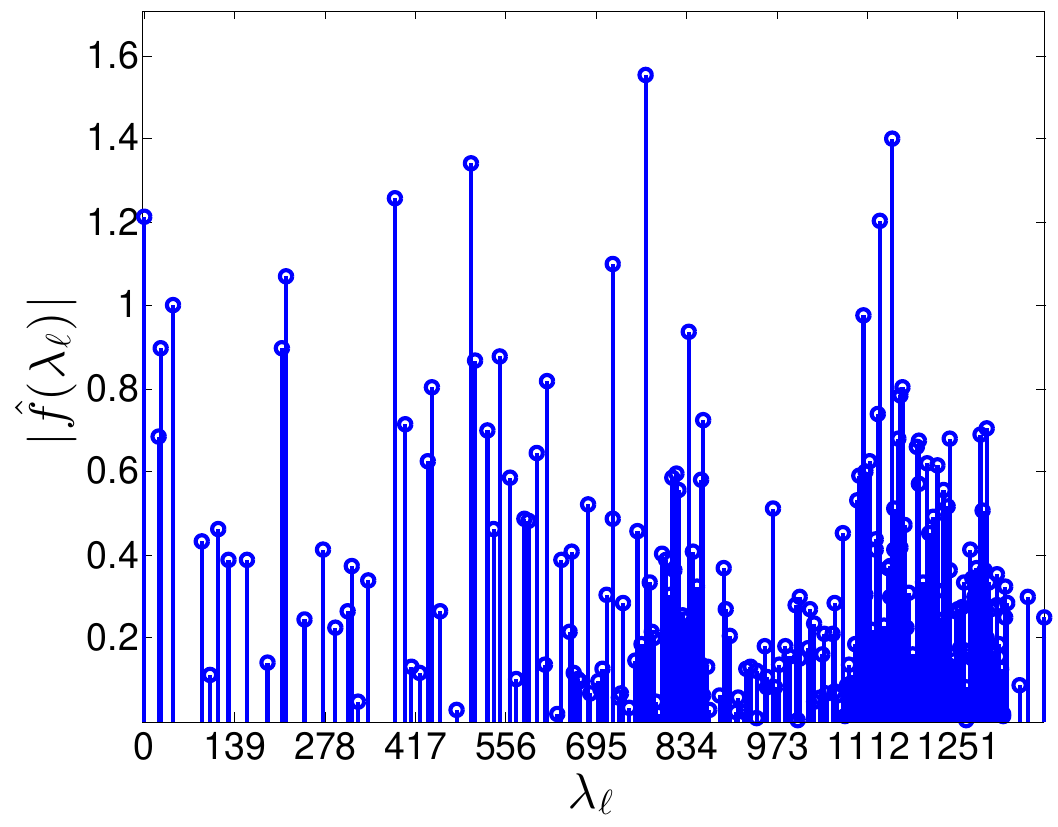}}
\centerline{\small{(b)}}
\end{minipage}
\hfill
\begin{minipage}[b]{.24 \linewidth}
   \centering
            \centerline{\includegraphics[width=\linewidth]{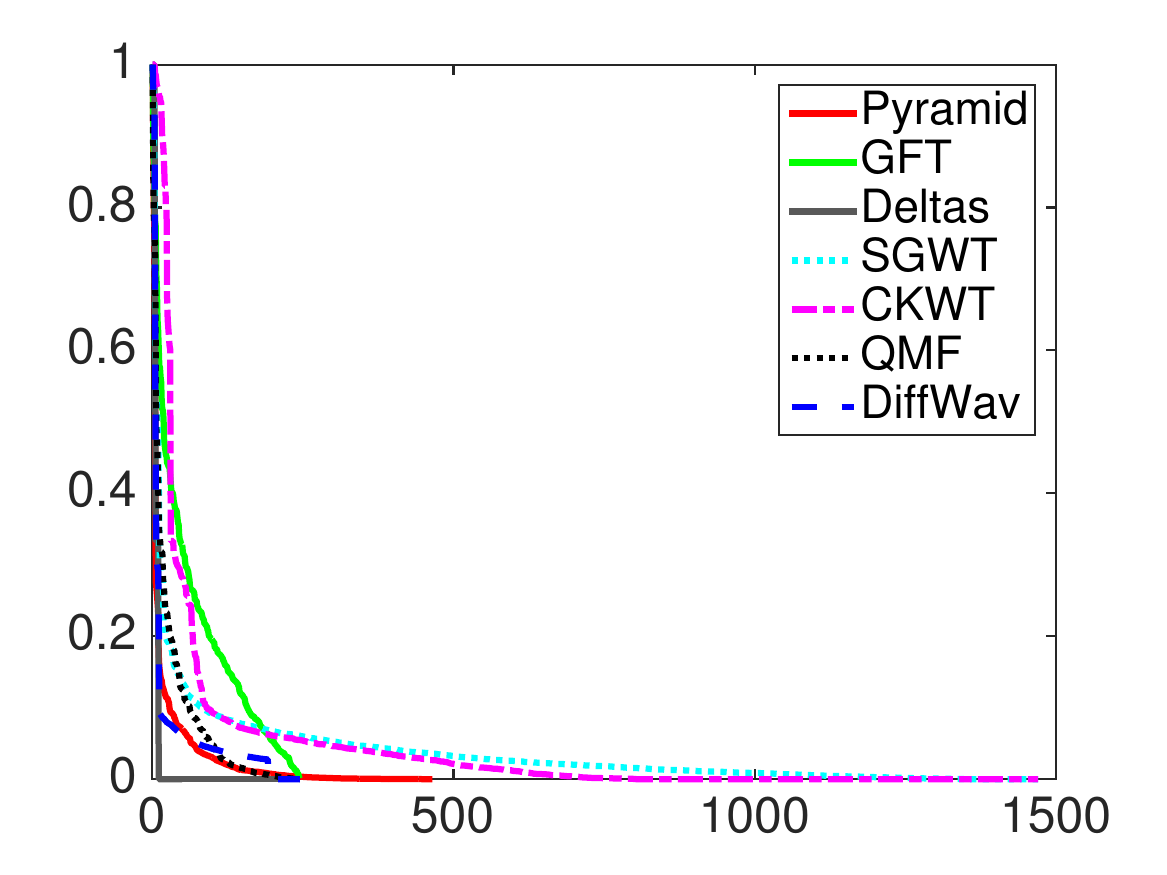}}
\centerline{\small{(c)}}
\end{minipage} 
\hfill
\begin{minipage}[b]{.24 \linewidth}
   \centering
         \centerline{\includegraphics[width=\linewidth]{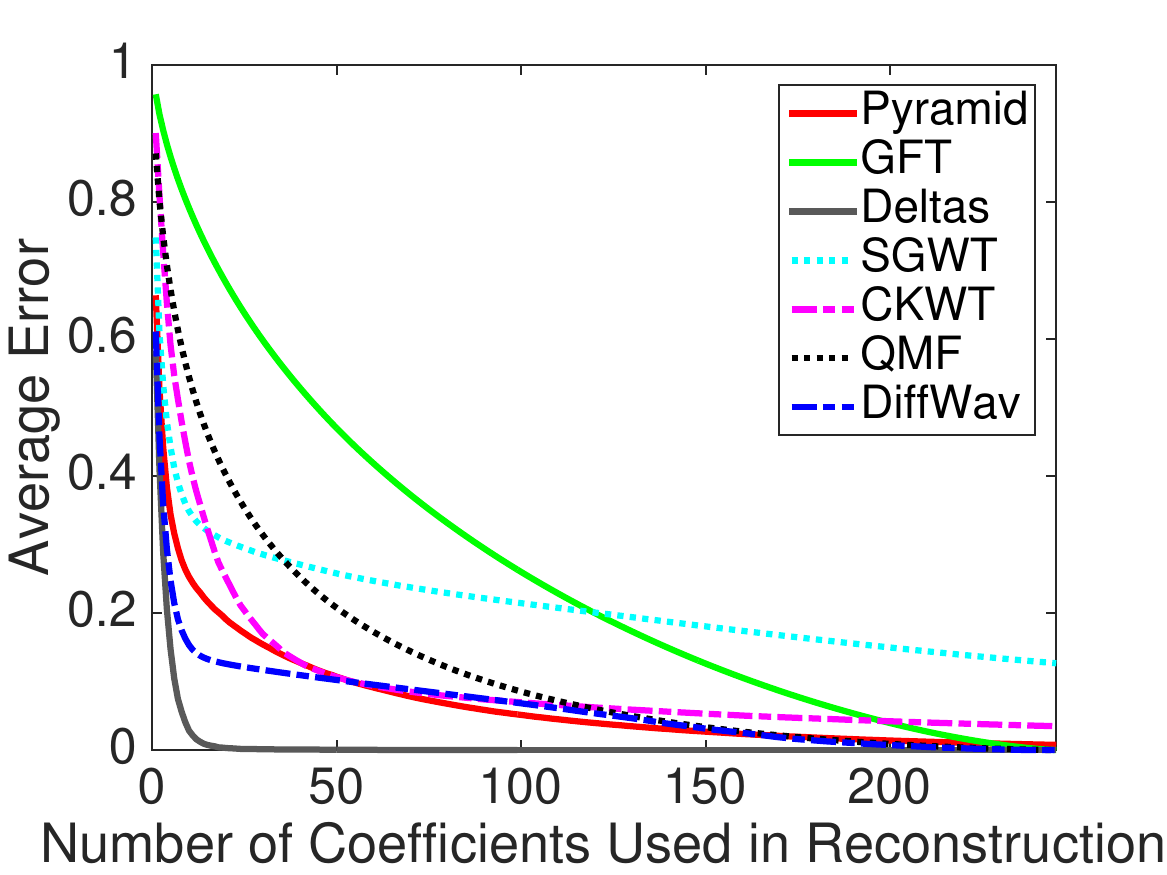}}
\centerline{\small{(d)}}
\end{minipage} 
\hfill
\caption {Compression example III. (a) A signal from the Flickr data set of \cite{Dong2013,thanou_TSP}. (b) The same signal in the graph spectral domain. (c) The normalized sorted magnitudes of the transform coefficients of the same signal. (d) The reconstruction errors are averaged over 500 signals from the data set.}
  \label{Fig:compression3}
\end{figure*}

\subsubsection{Compression}

Next, we repeat the compression experiment of Figure \ref{Fig:bunny_compressions} on three different types of graph signals. First, we consider a piecewise-smooth signal with discontinuities. We compose the signal in Figure \ref{Fig:compression}(a) on the 500 vertex random sensor network from Figure \ref{Fig:sparsification} by segmenting the graph into four strips and restricting two different polynomials of the coordinates to the different sections of the graph. The first polynomial $0.5-2\bar{x}$ is restricted to the first and third diagonal strips, counting from the upper right, and the second polynomial $0.5+\bar{x}^2+\bar{y}^2$ is restricted to the second and fourth strips (the latter of which is the lower left corner), where $(\bar{x},\bar{y})$ are the physical coordinates of each vertex. We sum the two components to form the signal.

The second signal, shown in Figure \ref{Fig:compression2}(a), is from \cite[Fig. 16(c)]{shuman_ACHA_2013}. It is formed by generating four different random signals on the graph, filtering them with different bandpass filters, restricting each of the bandpass signals to different regions of the graph, and then summing them. The first three regions are 4-hop neighborhoods of different center vertices, and the last region contains all vertices in the graph that are not in one of the first three regions. We can see for example that the smooth component in the center of the graph was filtered by a lowpass kernel in the graph spectral domain. 

The third class of signals we examine are from the Flickr data set, used previously in \cite{Dong2013} and \cite{thanou_TSP}. Each signal in the data set represents the number of distinct Flickr users that took photos in 10 x 10 meter regions around Trafalgar Square in London on a specific day. Like \cite{thanou_TSP}, we use a subgraph with 245 vertices, each one representing the centroid of a given 10 x 10 meter region. 
Note that while the signals from Figures \ref{Fig:compression}(a) and \ref{Fig:compression2}(a) are sparser in the graph spectral domain than the vertex domain, this Flickr data set contains signals that are much sparser in the vertex domain than the graph spectral domain. Figures \ref{Fig:compression3}(a) and \ref{Fig:compression3}(b) represent one example signal from the database in both the vertex and graph spectral domains. 

For all three compression experiments in this section, we use a three-level pyramid transform, take $\epsilon=0.005$ to form the regularized Laplacian for interpolation, and take the graph spectral filters to be lowpass Meyer kernels $\hat{h}^{(j)}(\lambda_{\l})$ whose values are equal to 1 on the interval $[0,\frac{\lambda_{\max}^{(j)}}{2}]$, smoothly decay to $\frac{\sqrt{2}}{2}$ at $ \frac{3\lambda_{\max}^{(j)}}{4}$, and reach 0 by $\lambda_{\max}^{(j)},$ where the superscript $j$ denotes the level of the pyramid. For comparison, we use (i) the graph Fourier transform (GFT); (ii) the basis of Kronecker deltas (i.e., compress the signal values directly); (iii) the spectral graph wavelet transform (SGWT) \cite{sgwt} with five wavelet scales plus the scaling functions for a total redundancy of 6; (iv) the graph-QMF filter bank transform (QMF) \cite{narang_bipartite_prod};\footnote{Note that the graph-QMF transform only uses the adjacency matrix structure, and not the weights of the graphs. The MATLAB code for the graph-QMF filter bank is publicly available at \url{http://biron.usc.edu/wiki/index.php/Graph\_Filterbanks}} 
(v) the spatial graph wavelets (CKWT) \cite{Crovella2003} with wavelet functions based on the renormalized one-sided Mexican hat wavelet, also with five wavelet scales and concatenated with the dictionary of Kronecker deltas (scale 0 dictionary atoms)\footnote{The spatial graph wavelets all have mean zero, so we add the dictionary of deltas to ensure that the dictionary analysis operator is full rank.} for a total redundancy of 6; and (vi) the diffusion wavelets of \cite{diffusion_wavelets}.\footnote{The MATLAB code for the diffusion wavelets is publicly available at \url{http://www.math.duke.edu/\textasciitilde mauro/code.html}}

Figures \ref{Fig:compression}(c), \ref{Fig:compression2}(c), and \ref{Fig:compression3}(c) show the decay of the magnitudes of the transform coefficients for the signals shown in Figures \ref{Fig:compression}(a), \ref{Fig:compression2}(a), and \ref{Fig:compression3}(a), respectively. In these plots, for each transform and signal pair, we normalize the sorted magnitudes by dividing every transform coefficient of that pair by the largest magnitude coefficient of that pair. For this compression experiment, we keep the $M$ coefficients with the largest magnitudes, set the rest to 0, and reconstruct the signal from the thresholded transform coefficients.
For a fair comparison, for all transforms, we perform least squares reconstruction by applying the pseudoinverse of the analysis operator to the thresholded coefficients.  In Figures \ref{Fig:compression}(d), \ref{Fig:compression2}(d), and \ref{Fig:compression3}(d), we show the reconstruction errors $\frac{||\mathbf{f}_{\textrm{reconstruction}}-\mathbf{f}||_2}{||\mathbf{f}||_2}$ for varying values of $M$, the number of coefficients used in the reconstruction. Note that the errors in \ref{Fig:compression3}(d) are averaged over 500 different signals from the data set.  We see that our proposed pyramid transform works reasonably well for compression for all three signal models.  One could also analyze the impact of using more sophisticated reconstruction methods, such as LASSO, matching pursuit, basis pursuit, and others (see, e.g., \cite{elad_book} for a survey of such methods).

\section{Implementation Approximations and Computational Complexity} \label{Se:approximations}
In their seminal paper on the Laplacian pyramid, Burt and Adelson \cite{burt_adelson} write, ``The coding scheme outlined above will be practical only if the required filtering computations can be performed with an efficient algorithm.'' In the same spirit, 
we mention 
here some approximation techniques to improve the computational efficiency of the proposed transform. 

As described analytically, the transform relies on repeated eigendecompositions of the graph Laplacian for filtering, downsampling, and graph reduction. However, for large graphs (with tens of thousands or more vertices), computing full eigendecompositions is computationally expensive and may not be possible. This is because general-purpose routines for computing the full eigendecomposition 
have computational complexities of $O(N^3)$. 
Accordingly, two common themes running through this section are (i) to avoid computing full eigendecompositions, 
and (ii) to take advantage of (and preserve) the sparsity of the graph Laplacian whenever possible.

\subsection{Power Method for Computing the Largest Eigenvector} \label{Se:power}
To compute the largest eigenvector of an $N \times N$ graph Laplacian, we can use the \emph{power method}, which generates a sequence of vectors $\left\{\mathbf{x}^{(k)}\right\}_{k=0,1,\ldots}$ through the following recursion:
\begin{align}\label{Eq:power_method}
{\mathbf{x}}^{(k)}=\frac{\L{\mathbf{x}}^{(k-1)}}{\norm{\L{\mathbf{x}}^{(k-1)}}_2}.
\end{align}
If $\lambda_{\max}>\lambda_{N-2}$ and $\langle \mathbf{x}^{(0)}, \mathbf{u}_{\max} \rangle \neq 0$, then the sequence of vectors $\left\{{\mathbf{x}}^{(k)}\right\}_{k=0,1,\ldots}$ generated by the power method converges to $\mathbf{u}_{\max}$ \cite[Chapter 8.2.1]{golub}. Convergence is faster the smaller the ratio $\frac{\lambda_{N-2}}{\lambda_{\max}}$ is, and it may be accelerated by subtracting a multiple of the identity matrix from $\L$ to shift the spectrum \cite[Chapter 4.1]{watkins}. The computational cost of \eqref{Eq:power_method} is dominated by the ${\cal O}(\card{\E})$ matrix-vector multiplication by the matrix $\L$.
In sparse graphs, $\card{\E}$ may grow linearly with $N$, making this an efficient method to compute the largest eigenvector. The condition $\langle \mathbf{x}^{(0)}, \mathbf{u}_{\max} \rangle \neq 0$ is satisfied by almost any $\mathbf{x}^{(0)}$ chosen at random \cite[Chapter 5.1]{watkins}. If the largest eigenvalue is not simple, the power method converges to a vector in the largest eigenspace, but this vector is not unique. %
For downsampling purposes, we only need to compute enough iterations to approximate the polarities of the largest eigenvector components, as opposed to their precise values.

\subsection{Efficient Implementation of Kron Reduction and Storage of Reduced Graph Information}

Full computation of the Kron-reduced Laplacian $\K\left(\L,\V_1\right)$ in \eqref{Eq:kron_def}  can be computationally expensive (${\cal O}(N^3)$ with naive methods that do not take advantage of the structure of $\L_{\V_1^c,\V_1^c}$).
One option is to not actually compute the entire reduced graph Laplacian, but rather to
form an operator to do matrix-vector multiplication with the matrix being the Kron-reduced Laplacian (see, e.g., \cite{saad} or the function MatCreateSchurComplement in PETSc \cite{petsc}). The drawback of doing so is that we cannot plot reduced graphs and coarser approximations as in Figure \ref{Fig:lp_example};
 however, such a method is well-suited for situations such as compression or regularization where we are only interested in computing transform coefficients.

In some cases, we may actually want to compute and store all of the graph Laplacians $\L^{(0)}, \L^{(1)},\ldots, \L^{(J-1)}$ as part of a multilevel transform. If we are repeatedly applying the transform to different signals on the same graph (e.g., a fixed transportation network), then the overhead of computing and storing this sequence of reduced graph Laplacians is shared across all of the applications of the transform. On the other hand, if the graph is completely different for every graph signal (e.g., some graph-based image processing methods), then efficient storage and reconstruction of the graph Laplacians is an important computational consideration. In either case, because the Schur complement is often dense even if $\L$ is sparse, it can also be computationally expensive to store $L^{Kron-reduced}$ for large graphs. The addition of a sparsification step after each Kron reduction reduces the storage requirements, as well as the complexity of subsequent computations. One interesting open question is whether we can avoid storing the entire sequence of Laplacians, but rather store the coarsest Laplacian and some additional side information, and still be able to recover the entire sequence in reverse order.

\subsection{Approximate Graph Spectral Filtering} \label{Se:chebyshev}
The Chebyshev polynomial approximation method, which is introduced in \cite{sgwt} and discussed further in \cite{shuman_DCOSS_2011}, can be used to speed up graph spectral filtering operations. 
That method computes an approximately filtered signal $\tilde{\mathbf{H}}\mathbf{f}$ 
through repeated matrix-vector multiplication by the matrix $\L$ at a computational cost of ${\cal O}({K}\card{\E})$, where $K$ is the order of the Chebyshev polynomial approximation. Therefore, for a sparse graph, it is far more efficient to compute $\tilde{\mathbf{H}}\mathbf{f}$ than ${\mathbf{H}}\mathbf{f}$.

\subsection{Computational Complexity} \label{Se:complexity}

To summarize, we can break the computational complexity down into two parts: the first for computing the multiresolution of graphs and the second for analyzing signals on that multiresolution of graphs with the filtering and interpolation operations. As mentioned above, the overhead of computing and storing the multiresolution of graphs is often shared across applications of the transform to many different signals on the same graph.

The bottleneck for the signal analysis portion of the multiscale pyramid transform we proposed in Section \ref{Se:pyramid} is the interpolation step \eqref{Eq:interp_kron}. Recent advances in 
solving such sparse, symmetric, diagonally dominant systems of equations
provide methods to reduce the computational cost  
from ${\cal O}(\card{\E}N)$ down to 
 ${\cal O}(\card{\E}\log^2 N)$, or even faster \cite{vishnoi,livne,koutis,spielman_web_page}. These advances also enable the spectral sparsification step to be done in nearly linear time with respect to the number of edges in the original graph.
To compute the full graph multiresolution for visualization, the bottleneck is the Kron reduction step \eqref{Eq:kron_def}. Here, we can again take advantage of methods for solving symmetric diagonally dominant systems of linear equations in nearly linear time in the number of edges. For example, we can use the LAMG solver of \cite{livne} to compute $\L_{\V_1^c,\V_1^c}^{-1}\L_{\V_1^c,i}$, for each $i$ in $\V_1$. Both the time to setup the problem (create a sequence of coarse-grained Laplacians) and the time to solve the problem are empirically shown in \cite{livne} to be linear in the number of edges. Since we have to solve the problem for each column of $\L_{\V_1^c,\V_1}$, the overall computational cost of the Kron reduction should be on the order ${\cal O}(|{\cal E}|N)$, but at least we only need to perform the setup portion once for $\L_{\V_1^c,\V_1^c}$.

\section{Conclusion}

Our main contribution in this work has been to present a new framework for a multiscale pyramid transform for graph signals that is centered around the four fundamental operations of graph downsampling, graph reduction, graph spectral filtering, and interpolation of graph signals. Our framework is modular, so we can easily substitute different choices for these fundamental operations (e.g., choose a different graph downsampling operator in the Laplacian pyramid of Figure \ref{Fig:lp}). We hope this framework inspires new choices for and analyses of these fundamental operations. For example, since the presentation of an initial version of our work here, \cite{nguyen} has presented a new method for graph downsampling in the context of signal processing, and  \cite{liu_coarsening} has presented a new method for graph reduction that preserves the eigenvalues from one graph to the next in the graph multiresolution. 

The interplay between the four operations is important and less well understood in the context of graph signal processing. Further empirical and theoretical investigations into how the joint choice of these operations affects properties such as the joint localization of the resulting dictionary atoms in the vertex-frequency space, as well as the sparsity of transform coefficients for different classes of graph signals is an important line of future work.  Another natural offshoot of this work would be to combine these fundamental graph signal processing operations in different manners to generate other multiscale transforms such as critically sampled filter banks and lifting transforms.

\balance


\bibliographystyle{IEEEtran}
\bibliography{multiscale}

\end{document}